\documentclass[preprint]{aastex}

\newcommand{\lbol}{L$_{bol}$}
\newcommand{\llyc}{L$_{LyC}$}
\newcommand{\lk}{L$_{K}$}
\newcommand{\nsn}{$\nu_{SN}$}

\newcommand{\htwo}{\mbox{H$_2$}}
\newcommand{\CO}{$^{12}$CO}

\newcommand{\COz}{$^{12}$CO 2-1 }
\newcommand{\arasa}[2]{Ann. Rev. Astron. Astrophys., #1, #2.}

\newcommand{\apjj}[2]{Ap. J., #1, #2.}
\newcommand{\apjjs}[2]{Ap. J. Supp., #1, #2.}
\newcommand{\apjjl}[2]{Ap. J. (Letters), #1, #2.}

\newcommand{\asa}[2]{Astron. Astrophys., #1, #2.}

\newcommand{\asas}[2]{Astron. Astrophys. Suppl., #1, #2.}
\newcommand{\mn}[2]{M.N.R.A.S., #1, #2.}

\newcommand{\ajj}[2]{A. J., #1, #2.}

\newcommand{\vol}[1]{1}

\newcommand{\solar}{L$_{\odot}$\ }

\newcommand{\solm}{M$_{\odot}$}

\newcommand{\rf}{\par\noindent\hangindent 15pt {}}

\shorttitle{The Nuclear Stellar Cluster in the Seyfert 1 NGC~3227}
\shortauthors{Schinnerer, Eckart \& Tacconi}

\begin{document}

\title{The Nuclear Stellar Cluster in the Seyfert~1 Galaxy NGC~3227:
       High Angular Resolution NIR Imaging and Spectroscopy}

\author{E. Schinnerer}
\affil{Astronomy Department, California Institute of Technology,
       Pasadena, CA 91125}
\affil{MPI f\"ur extraterrestrische Physik, Giessenbachstrasse, 85748
       Garching, Germany}
\email{es@astro.caltech.edu}

\author{A. Eckart}
\affil{I. Physikalisches Institut, Universit\"at zu K\"oln,
       Z\"ulpicherstra\ss e 77, 50937 K\"oln, Germany}
\affil{MPI f\"ur extraterrestrische Physik, Giessenbachstrasse, 85748
       Garching, Germany}
\email{eckart@zeus.ph1.uni-koeln.de}

\and

\author{L.J. Tacconi}
\affil{MPI f\"ur extraterrestrische Physik, Giessenbachstrasse, 85748
       Garching, Germany}
\email{linda@mpe.mpg.de}


\begin{abstract}
Near-Infrared high angular resolution speckle imaging and imaging
spectroscopy of the nuclear region ($\sim$ 10'' $\approx$ 840~pc)
of the Seyfert~1 galaxy NGC~3227
are presented. The images reveal an unresolved nuclear source in the $K$ 
band in addition to a nuclear stellar cluster which is slightly resolved 
in the $J$ and $H$ band. The contribution of this stellar cluster to
the NIR continuum is increasing from the $K$ to the $J$ band.
The stellar absorption lines are extended compared to the neighboring
continuum suggesting a nuclear stellar cluster size of $\sim$ 70~pc FWHM.
Analysis of the stellar absorption lines suggests that the stars are
contributing about 65\% (40\%) of the total continuum emission in the
$H$ ($K$) band in a 3.6'' aperture. The dominant stellar type are cool 
M type stars.
Population synthesis in conjunction with NIR spectral synthesis
indicates that the age of the mapped nuclear stellar cluster is in the
range of 25 to 50~Myr when red supergiants contribute most to the NIR light.
This is supported by published optical
data on the Mg I $b$ line and the CaII triplet. Although a higher age of 
$\sim$ 0.5 Gyr where AGB stars dominate the NIR light can not be excluded,
the observed parameters are at the limit of those expected for a
cluster dominated by AGB stars. However, in either case the
resolved stellar cluster contributes only about $\sim$ 15 \% of the
total dynamical mass in the inner 300~pc. This immediately implies at least
another much older stellar population which contributes to the mass
but not the NIR luminosity. 
Pure constant star formation over the last 10 Gyr can be excluded
based on the observational fact that in such a scenario the 
total observed (spatially unresolved and spectrally resolved) Br$\gamma$ 
flux would be of stellar origin which is spatially extended.
Therefore, at least two star formation/starburst events took place in
the nucleus of NGC~3227. Since such sequences in the nuclear star
formation history are also observed in the nuclei of other galaxies a
link between the activity of the star formation and the AGN itself
seems likely.
\end{abstract}


\section{\label{ww01}INTRODUCTION}

The nuclear stellar properties of Seyfert galaxies are of great
interest, since there might be a connection between the nuclear star
formation and the AGN activity (Norman \& Scoville 1988). Current
scenarios for fueling the nuclear region involve nuclear bars
(e.g. Shlosman, Frank \& Begelman 1989). However, recent high-angular
resolution observations of Seyfert galaxies do not find enough
evidence for this model (Regan \& Mulchaey 1999, Malkan, Gorjian \&
Tam 1998). The nuclear star formation activity and the
AGN both use the same fuel -- the circumnuclear molecular gas. Detailed
analysis of the nuclear stellar properties of the two Seyfert galaxies
Circinus (Maiolino et al. 1998) and NGC~1068 (Thatte et al. 1997)
have shown that the nuclear stellar clusters have age of $\sim$ 10$^8$
yr. This suggests that at least some of the molecular gas reaching the
inner few 100~pc is converted into stars rather than being funneled directly
to the AGN. In addition there might be an interplay between the star
formation and the AGN classification (Ohsuga \& Umemura 1999).
\\
\\
High angular resolution observations in the near-infrared (NIR) of the
nuclear region of active galaxies are well-suited to investigate this
problem via a quantitative analysis of the contributions of the AGN and the
stars to the nuclear NIR continuum. The NIR offers two advantages over 
the optical wavelength
ranges: (1) the extinction affecting the stellar light is much smaller
(about 1/10 in the $K$ band compared to the $V$ band) and (2) the
overall stellar contribution to the continuum emission is larger due
to the spectral energy distribution (SED) of the AGN and the stars
(e.g. Barvainis 1989). This allows for a more detailed analysis of the
properties of the stellar contribution to the nuclear light. The
combination of imaging spectroscopy with high angular resolution
imaging is ideal to disentangle the contribution of the AGN from those
of the stars using spectral and spatial information. The utilization
of population synthesis and NIR spectral synthesis models allows to
analyze the star formation history of the nuclear stellar component.
NGC~3227, a nearby Seyfert~1 galaxy, was chosen for such a detailed
study, since complementary data from other wavelengths are already 
available in the literature.
\\
\\
NGC3227 (Arp~94b; type SAB pec, de Vaucouleours et al. 1993) is a Seyfert 1 
galaxy located at a distance of 17.3~Mpc
(group distance; Garcia 1993; 1'' $\sim$ 84~pc) and interacting with
its elliptical companion NGC~3226. 
Rubin \& Ford (1968) studied the system NGC~3226/7 for the first time in the 
optical. In NGC~3227 they found indications for a nuclear outflow 
as well as a spiral arm that stretches toward the close elliptical companion.
The nucleus of NGC~3227 exhibits clear signs of Seyfert~1
activity. The Narrow Line Region (NLR) is extended towards the northeast 
(Mundell et al. 1992a, Schmitt \& Kinney 1996). The Broad Line Region
(BLR) clearly shows variations in the 
optical continuum and line emission (Salamanca et al. (1994).
Mundell et al. (1995b) detected a double radio source with a separation
of 0.4'' at a PA of $\sim$10$^o$.
The cold molecular gas traced by its \CO ~line emission shows a
circumnuclear ring with a diameter of $\sim$ 3'' plus weaker
emission at the dynamical center. In the inner
arcsecond strong non-circular motions are observed. Modeling of the
kinematics indicates that a warped thin molecular gas disk can
explain the observed kinematics quite well (Schinnerer, Eckart \&
Tacconi 1999, 2000). 
Optical and NIR observations (Mulchaey, Regan, Kundu 1997, De
Robertis et al. 1998) indicate that the galaxy has a bar with R
$\approx$ 6.7 - 8.4~kpc and a position angle of about -20$^o$ relative 
to the major kinematic axis. 
\\
\\
In this paper we present the first quantitative analysis of the AGN
and stellar contributions to the nuclear NIR continuum of NGC~3227
using high angular NIR imaging and integral field spectroscopy. We
describe our observations in \S \ref{ww02}, and 
discuss the results from the integral field spectroscopy in \S \ref{ww04}. 
The nuclear stellar content is derived in \S \ref{xxs55}.
We present an analysis of
the high angular resolution imaging data in \S \ref{ww03}. 
Population synthesis and
NIR spectral synthesis models are used in \S \ref{xxs7} to analyze the nuclear 
star formation history of NGC~3227. A description and discussion of
these models is given in the appendix. A brief summary is presented in \S
\ref{ww08}.

\section{\label{ww02}OBSERVATIONS}

\subsection{NIR imaging spectroscopic data}

NGC~3227 was observed with the MPE integral field spectrograph 3D in
the $H$ and $K$ band during two observing runs in January 1995 at the
3.5~m telescope on Calar Alto, Spain, and in December 1995/January 1996
at the William-Herschel-Telescope (WHT) at La Palma, Spain (Table
\ref{xxt01}). The
spectral resolution was $R=\frac{\lambda}{\triangle \lambda} \approx$
750 for the $K$ band and $\approx$ 1000 for the $H$ band.
3D (Weitzel et al. 1996) was used together with the tip-tilt corrector
ROGUE (Thatte et al. 1995). 3D obtains simultaneous images and
spectra over an 8'' $\times$ 
8'' field. This is done using an image slicer which
rearranges the two-dimensional focal plane onto a
long slit. A grism is used for  dispersion and the
 spectra are then detected on a
NICMOS~3 array. A detailed description of the instrument and the data reduction
is given by Weitzel et al. (1996). 
\\
The data reduction procedure converts
each two dimensional image into a three dimensional data cube with two
spatial axes and one spectral axis. The data cubes are co-added and centered
on the continuum peak. All images are dark-current and sky-background
subtracted, corrected for dead and hot pixels, and spatially and
spectrally flat-fielded. To correct the effects of the Earth's
atmosphere to the $H$ and $K$ band spectrum, a standard star was observed.
This standard spectrum was first divided by a template spectrum of the same
spectral type (Kleinmann \& Hall 1986) in order to remove stellar
features. In the $H$ band standard star observations could not be
obtained for one night. We therefore used the model atmosphere from ATRAN 
for correction. Since the correction is only reliable for the central 
wavelength range (1.5196 $\mu$m - 1.7248 $\mu$m),
we discarded the remaining parts of the band.
The effect due to different zenith distance of source and
standard star was minimized using the ATRAN atmospheric model (Lord
1992), mainly to correct for the different atmospheric absorption. The
source data were then divided by the resulting atmospheric transmission 
spectrum.
\\
The flux calibration was done in a 4.6'' aperture using the flux
measurement from our high resolution imaging SHARP data. From a
comparison with the reference star and with the SHARP data
we estimate that
the achieved spatial resolution is $\sim$ 1.6'' FWHM and
$\sim$ 1.3'' FWHM for the $H$ and $K$ band, respectively. 
The absolute flux
calibration uncertainties are $\sim$ 10\%. The uncertainties
for the line fluxes given in Tables \ref{xxt07} and \ref{xxt08}
include the noise in the line map as well as
uncertainties due to the choice of the baseline and absolute flux
calibration uncertainties. The total uncertainty amounts to about 13\%. For the lines close to
2.00 $\mu$m the uncertainties are larger due to the larger uncertainties
induced by the low atmospheric transmission. 
Due to different implementations of the $K$ band grism in the two
observing runs the coverage of the $K$ band differed slightly
resulting in lower S/N at the beginning and the end of the spectral band.

\subsection{\label{xxsh}NIR speckle data}

NGC~3227 was observed in the $J$, $H$ and $K$ band with the MPE speckle 
camera SHARP1 mounted to the ESO New Technology Telescope (NTT) at La
Silla, Chile, on four nights in June 1996 (Table \ref{xxt01}). 
SHARP1 has a 12.8'' $\times$ 12.8'' field of view with a
pixel scale of 0.05''/pixel.
The star 35~Leo served as a flux calibrator and a PSF reference.
The images were sky-background subtracted,
corrected for dead pixels and flat-fielded, and then
co-added after re-centering with respect to each other using a simple
shift and add (SSA) algorithm (Christou 1991). 
The nucleus of NGC~3227 was alternately positioned
in one of the four quadrants of the NICMOS3 detector. The resulting
images at different positions were mosaiced to obtain a larger FOV. 
The integration times per frame ranged from 0.5s ($H$, $K$ band) to
1.5s ($J$ band) for NGC~3227 and from 0.25s ($K$ band) to 0.5s ($H$,
$J$ band) for 35 Leo. At each position about 300 frames were taken
leading to an average integration time of $\sim$ 3 minutes (per 
SSA image) for both, NGC~3227 and 35~Leo.
\\
\\
\underline{\it The stability of the PSF:}
For a reliable estimate of the nuclear size of NGC~3227, good 
knowledge of the stability of the PSF is necessary. The
PSF standard star is located about 5$^o$ away from NGC~3227 which is
sufficiently close to monitor the same atmospheric conditions. 
The standard star 35 Leo was observed at least ones per band,
either at the beginning or end of the source observations each night.
The total duration of the observation per band and night ranged from
20 to 75 minutes. To verify that the seeing was stable, we measured
the FWHM in the individual SSA images of NGC~3227 and 35~Leo for each
night. The FWHM within each band was found to be stable and the observed
differences between NGC~3227 and 35~Leo were similar for all the nights.
Therefore the seeing was stable during the observations and better than 1'' in
the visible (see Table \ref{xxt01}): 
\\
Uncertainties in the PSF due to different frame integration times for NGC~3227
and 35~Leo can be excluded: The
integration time per single frame ranged from 0.25s to 1.5s which is
still short enough to obtain only a small number of speckles per
frame, since the atmospheric coherence time in the $K$ band lies between
30ms to 100ms (Hofmann et al. 1995). Different integration times (per
frame) for NGC~3227 and the star (up to three times longer) might be
considered as a cause for instability. However, comparison of our $K$
band data with the same integration time used for NGC~3227 and the
standard star, and different integration times used for both showed no
difference in the obtained results. The seeing is expected to be
better at longer wavelengths, since the atmospheric variations are
slower and the monitoring is easier (see Table \ref{xxt01}). 
All the above tests make us confident that the observations of 35~Leo gave
a stable and reliable PSF for the NGC~3227 data. 
\\
\\
The relative uncertainties in the flux calibration of the individual nights,
and NIR bands amount to (5 - 8)\%.
Including systematic uncertainties (differences between 
color systems and filter curves) we estimate the total uncertainty for the flux
calibration to be $\sim$ 10\%.
Comparison of the calibrated SSA mosaics (Table \ref{xxt02})
and calibrated direct imaging
data with a larger field of view (McAlary et al. 1983, Kotilainen et
al. 1992) yields agreement to within 0.15$^{mag}$.
\\
The achieved angular resolution is 0.55'' FWHM in the $J$ band and
0.35'' FWHM in the $H$ and $K$ band as measured from the reference
star. 

\subsection{HST $V$ band image}

The HST F606W image of NGC~3227 (P.I. Malkan) is saturated in the
nucleus. We extracted the PC1 chip part with a pixel scale of
0.0455''/pixel. 
The image was corrected for dead pixel and remaining
'hot' pixel were removed by hand. The resolution in the HST image is 0.13'' as
obtained from a Gaussian fit to the northern knot (see section \ref{xxcr}, 
Fig.  \ref{xxa01}). 
The correct center position was
obtained by aligning the HST image with respect to the SHARP1 data. For
this purpose the data were smoothed to the SHARP1 pixel scale of 0.05'' and convolved to
the SHARP1 angular resolution of 0.35''. To minimize for effects from
the saturated nuclear pixels we radially interpolated the flux from
the neighboring pixels before convolving and then masked the
corresponding pixels in the 0.35'' image. To obtain 
the correct center the position of the northern knot and the position 
angle given in the
header were used. The data was calibrated using the values from McAlary et al.
(1983) in ring apertures, since the central pixels were saturated.

\section{\label{ww04}NIR IMAGING SPECTROSCOPY}

\begin{figure}[t!]
\begin{center}
\includegraphics[height=15.5cm,width=10.5cm,angle=-90.]{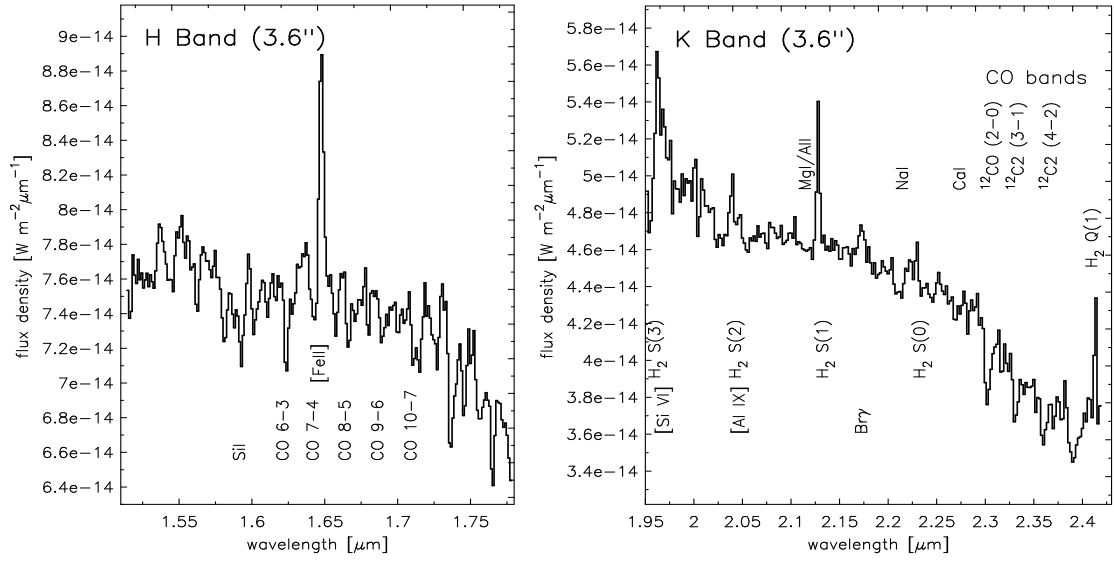}
\end{center}
\figcaption{\label{xxa06}
3D spectra of NGC~3227 in $H$ and $K$ band in a circular aperture of
3.6''. 
}
\end{figure}

The $H$ and $K$ band spectra (Fig. \ref{xxa06}) are fairly flat suggesting
effects of extinction, dust emission and maybe an AGN power 
law continuum may all contribute in addition to the stellar light.
This is consistent with the findings from the NIR colors (see section
\ref{xxse}).
\\
The prominent emission lines in the spectra are the [Fe~II] line at 
1.64$\mu$m in the
$H$ band, and \htwo ~lines -- especially the \htwo 1-0 S(1) line
at 2.12$\mu$m --, Br$\gamma$, and [Si VII] lines in the $K$ band. 
Also a number of stellar absorption lines of CO, OH, SiI, NaI and CaI
are present. We describe these spectral features in the following
sections. Our data agree well with line fluxes and equivalent widths
from the literature (Table \ref{xxt06}).
All line fluxes in this paper were derived from the line maps.

\subsection{\label{xxs4}BLR, CLR and NLR emission lines}

\begin{figure}[t!]
\begin{center}
\includegraphics[height=15.5cm,width=10.5cm,angle=-0.]{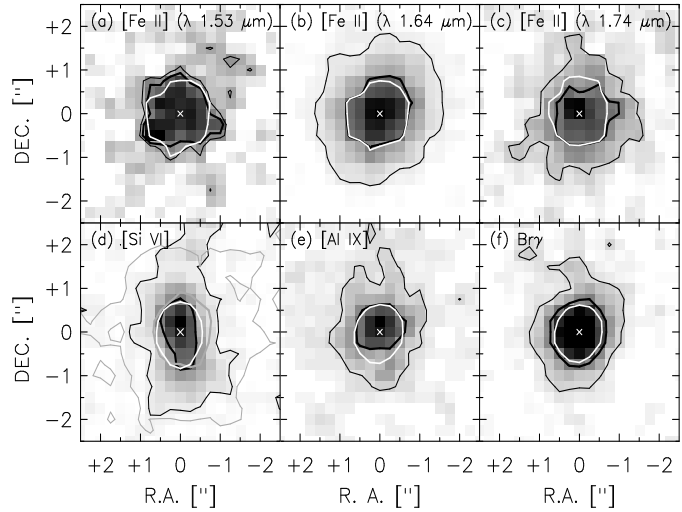}
\end{center}
\figcaption{\label{xxa07}
Maps of the AGN emission lines in the nucleus of NGC~3227.
The 50\%
contour lines of the line emission (fat black line) and the neighboring
continuum (fat white line) are shown. The thin black line indicates the
3$\sigma$ level in the maps. In the [Si VI] line image (d) the
corresponding contours of the neighboring \htwo 1-0 S(3) line are
shown as well to demonstrate the different spatial extend of both
lines (50\% contour line: fat gray line; 3$\sigma$ level: thin
gray line). For further discussion see text.
}
\end{figure}

In active galaxies emission lines arising at various distances from the central
engine can be sub-divided into three main classes. Permitted
emission lines with line widths of a few 10$^4$ km/s arise in the
BLR with a size of less 1~pc. Forbidden lines
with high ionization potential ($\chi >$ 100 eV) arise in the 'Coronal
Line Region' (CLR) of $\sim$ 100~pc. Permitted
and forbidden emission lines with moderate line widths from 
the NLR can extend over $\sim$ 1~kpc. Emission line maps and fluxes from
all three regions are presented in Fig. \ref{xxa07} and Table
\ref{xxt07}.
\\
\\
\underline{\it Lines from the BLR:} 
The Br$\gamma$ line emission in NGC~3227 is mostly unresolved
at our resolution of 1.3''. However, at the 10\% level of the peak
emission, extended flux towards the northeast of the nucleus is apparent. The
Br$\gamma$ line has a FWHM of $\sim$ 650 km/s. However, due to the triangular
line profile (also observed by Vanzi et al. 1998) and our
spectral resolution of $\sim$ 350 km/s, it is difficult to determine
the exact width. However, the Br$\gamma$ FWZI of $\sim$ 2400 km/s is 
in reasonable agreement with the 2900 km/s found for the variable component 
of the H$\alpha$ line (Salamanca et al. 1994) given the uncertainties.
\\
The peak of the extended Br$\gamma$ component coincides with the $K$ band continuum peak
indicating that the BLR is at the position of the continuum.
Comparison of the distribution of the molecular gas and its velocity
field with the $VJHK$ colors suggests that the $K$ band peak and the dynamical
center are identical within the uncertainties. This is in contradiction to
the result of Arribas \& Mediaville (1994) who found an
offset between the BLR and the dynamical center. Since parts of the
H$\alpha$ emission arise in the NLR (see below) and their proposed
dynamical center lies in a region with lower H$\alpha$ emission (see
Figure 5 of Arribas \& Mediaville 1994), the velocity
field of the H$\alpha$ line emission could be dominated by NLR 
kinematics rather than that of the gravitational potential. 
\\
\\
\underline{\it The NLR emission lines:}
The emitting gas in the NLR of active galaxies 
has low density (10$^2$ - 10$^6$) and a
large spatial extent (10~pc - 1~kpc) which is often cone-like (see
review by Wilson 1997). The main excitation mechanism in this region
is photo-ionization. The Br$\gamma$ emission in NGC~3227 is extended at a very low
flux level. The structure and orientation is in agreement with the
ionization cone as seen in the [O~III] line emission (Schmitt \&
Kinney 1996) and also in the H$\alpha$ map of Arribas \& Mediaville
(1994). The $H$ band [Fe~II] line emission is still unresolved probably due to
the fact that the S/N and spatial resolution is not as high as in the 
$K$ band data cube. The width of the [Fe~II] line is $\sim$ 650 km/s
FWHM typical for NLR lines and comparable to the observed line width of
the Br$\gamma$ line. This indicates that the Br$\gamma$ emission is
dominated by NLR gas emission.
\\
Simpson et al. (1996) suggested that the [Fe~II] emission in active
galaxies is mainly due to photo-ionization of material with normal ISM
abundances by the hard AGN continuum emission. However, a significant
fraction ($\sim$ 20\%) can still be shock excited. 
In NGC~3227 the observed $\frac{[Fe II]}{Br \gamma}$ ratio in a 3.6'' 
aperture is 2.9$\pm$0.7 lying
between the values found for NGC~1068 and NGC~7469 (Mouri et al. 1990). 
Excitation by SN remnants (Oliva et al. 1989, Lumsden \& Puxley 1995)
can be excluded, since in this scenario the [Fe II] and the \htwo ~line
emission should spatially coincide.
Therefore it seems highly likely that the [Fe II] line is arising
from the NLR gas, whereas the \htwo ~emission is probably
collisionally excited via shocks (Schinnerer, Eckart \& Bertoldi, in
prep.).
\\
\\
\underline{\it The CLR emission lines:}
Due to the high ionization potential ($\chi$ $>$ 100 eV) associated with CLR
emission lines high energetic processes are required. If the
excitation occurs via collisions, the gas temperature should be about
$T \sim 10^6$K similar to the corona of the sun. In the case of
photo-ionization via the hard AGN continuum, temperatures of only a few
10$^3$ - 10$^4$K are needed, however.
\\
In NGC~3227 the [Si VI] and marginally the [Al XI] emission lines are
detected and spatially unresolved at our resolution and sensitivity. 
Both lines are located in a wavelength range
that is affected by atmospheric absorption features and are close
to \htwo ~lines. The same baseline fit was used for the CLR lines and their
adjacent \htwo ~lines. Since the CLR lines are spatially unresolved relative
to their neighboring continuum whereas the
adjacent \htwo ~lines are spatially extended (Fig. \ref{xxa07}), we are,
therefore confident of their detection. However, due to the
different spatial extent of the CLR lines and the diluting \htwo
~lines it is not straight-forward to obtain a corrected CLR line flux.

\section{\label{xxs55}THE STELLAR ABSORPTION LINES}

\begin{figure}[t!]
\begin{center}
\includegraphics[height=15.5cm,width=10.5cm,angle=-0.]{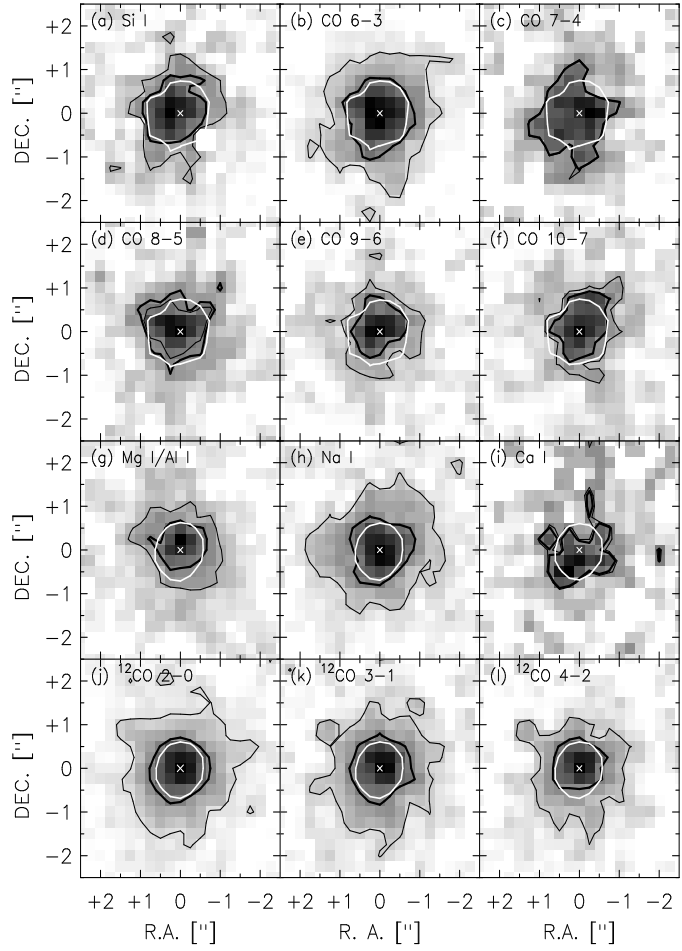}
\end{center}
\figcaption{\label{xxa08}
Maps of the extended stellar absorption lines in the nucleus of
NGC~3227. The 50\%
contour lines of the line absorption (fat black line) and the neighboring
continuum (fat gray line) are shown. The thin black line indicates the
3$\sigma$ level in the maps.
}
\end{figure}

Prominent stellar absorption lines detected in the $H$ and $K$ band come
from CO, OH, SiI and NaI and arise in the atmosphere of late-type
stars. Therefore a noticeable stellar contribution to the continuum is
expected. The stellar absorption lines are extended compared to the
neighboring continuum with FWHM of 1.5'' and 1.8'' in the $K$ and $H$ band, 
respectively. With a seeing of $\sim$ 1.3'' ($K$ band) this leads to a size 
of the nuclear stellar cluster of
0.8'' ($\approx$ 70~pc) after quadratic deconvolution. Line fluxes and
equivalent widths are summarized in Table \ref{xxt08}. The line maps
are shown in Fig. \ref{xxa08}. All analysis
is done in an aperture of 3.6'' which includes most of the stellar
line flux in the corresponding images. 
\\
The equivalent widths of SiI (1.58 $\mu$m), CO 6-3 (1.62 $\mu$m) and
CO 2-0 (2.29 $\mu$m) can be used to classify the dominating stellar
type allowing an estimate of the age of the luminous nuclear stellar cluster.
Late-type dwarfs as the main contributers can be excluded due to their
low luminosities.

\subsection{The spectral class from the SiI and CO 6-3 lines}

Origlia et al. (1993), Oliva et al. (1995), Origlia et al. (1997),
Oliva \& Origlia (1998), Origlia et al. (1999) have analyzed the
properties of the SiI (1.58 $\mu$m), CO 6-3 (1.62 $\mu$m) and
CO 2-0 (2.29 $\mu$m) lines, and have used the ratios and equivalent
widths (EWs) to classify stars as well as stellar populations and clusters. 
To obtain the equivalent width
$W_{\lambda}$ the neighboring continuum $f_{\lambda}$
is rectified and normalized to unity, 
the line is then integrated between the limits $\lambda_1$ and $\lambda_2$
(as given by Origlia et al. 1993):

\begin{equation}
W_{\lambda} = \int_{\lambda_1}^{\lambda_2} (1 - f_{\lambda}) d \lambda
\end{equation}

\begin{figure}[t!]
\begin{center}
\includegraphics[height=15.5cm,width=10.5cm,angle=-0.]{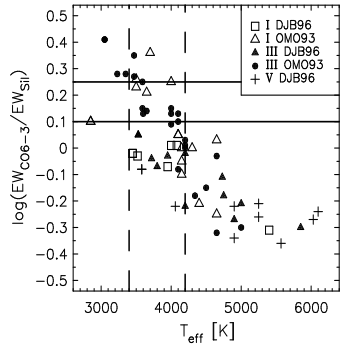}
\end{center}
\figcaption{\label{xxa09}
Diagnostic line diagram for NGC~3227. The observed range of the
$log(\frac{CO 6-3}{Si I})$ is indicated (fat solid lines). The resulting
range for the effective temperature is given by the broken lines.
(DJB96: Dallier, Boisson, Joly 1996; OMO93: Origlia, Moorwood, Oliva
1993) 
}
\end{figure}

\noindent
The SiI line is only present in stars later than type G4 before it
is diluted by the Br14 emission line. The equivalent width depends
only on the SiI abundance and therefore the metallicity.
The CO lines are observed in stars with $T_{eff}$ $<$ 4500 K
and depend on the effective temperature, the surface gravity,
the CO abundances (CO 6-3) and the velocity of the microturbulance (CO
2-0). 
\\
\\
We can estimate the metallicity from the shape of the CO 6-3 line complex,
since it also includes contributions from OH and Ca lines. Oliva
\& Origlia (1998) and Origlia et al. (1997) show synthesized spectra of this
line complex for different metallicities, effective temperatures, and
micro-turbulent velocities in the stellar atmospheres.
\\
In NGC~3227 the absorption associated with the CO molecule is significantly
stronger than that of the OH line. This indicates that the metallicity in
NGC~3227 is at least similar to that inferred for starburst galaxies (Origlia et al
1997) where metallicities of $[\frac{Fe}{H}]$ $\sim$ -0.5 to -1.0 are
observed. These calculations have used an upper limit for the effective
temperature of 3600~K for the red supergiants, so the metallicities are very 
likely even higher, since hotter stars need higher metallicities to explain the
observed line profiles. Therefore the metallicities inferred for
starburst galaxies should be regarded as lower limits (Oliva \& Origlia
1998). 
\\
The ratio of the equivalent widths of $\frac{CO~6-3}{SiI}$ can be used to
estimate the effective temperature of the dominant stellar type in
the nuclear stellar cluster of NGC~3227. Since both lines are close in
wavelength range, diluting effects such as extinction or non-stellar
continuum contributions can be neglected. For NGC~3227 
we find within the uncertainties an
effective temperature of (3200 - 4200)K for red giants or supergiants
suggesting a spectral class of K3 - M6 for giants and K2 - M4 for
supergiants (Fig. \ref{xxa09}).
Comparison of the measured equivalent widths (SiI: (1.2$\pm$0.2)\AA ~and CO
6-3: (1.9$\pm$0.1)\AA) to the expected widths (SiI: (4$\pm$1)\AA ~and
CO 6-3: (6$\pm$2)\AA) suggests a stellar contribution of late-type stars 
to the continuum of (30$\pm$10)\% in the $H$ band.

\subsection{Further restrictions on the spectral class from the OH line}

\begin{figure}[t!]
\begin{center}
\includegraphics[height=15.5cm,width=10.5cm,angle=-90.]{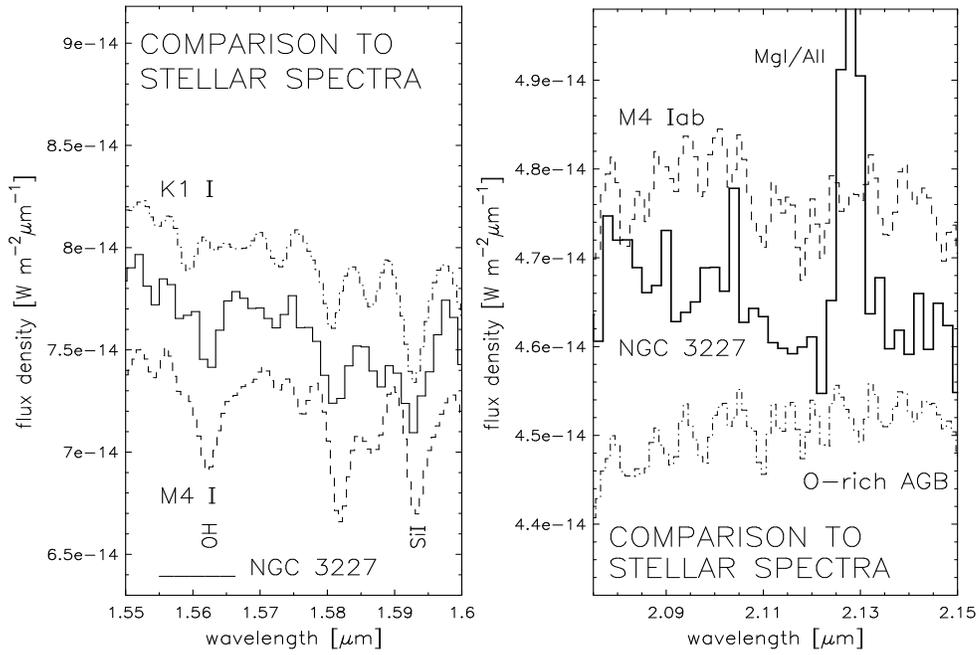}
\end{center}
\figcaption{\label{xxa10}
Comparison of the OH line in NGC~3227 to standard star spectra (K1 I:
93~Her, M4 I: HR 921; from Meyer et al. 1998). Comparison of the $K$ band
spectrum of NGC~3227 to spectra of an AGB star and a supergiant 
(S6+/le: $\chi$ Cyg, M3-4 Iab: SU Per;
from Wallace \& Hinkle 1997) scaled and continuum added according the
line depths.
}
\end{figure}

Meyer et al. (1998) have investigated the dependence of the effective temperature and
luminosity class on the $H$ band stellar absorption lines. They find that
the OH line at 1.689 $\mu$m is a good indicator for effective
temperatures below 4400 K. The OH absorption increases with decreasing
effective temperature. The line feature is clearly visible in spectra
of red giants and supergiants later than type M0. An additional OH
line at 1.542 $\mu$m shows a similar behavior. 
\\
In NGC~3227 a strong absorption line is present at 1.548 $\mu$m which
is in agreement with the wavelength of the red-shifted OH line (Fig.
\ref{xxa10}). This
indicates that the stellar light is dominated by cool M stars. The
effective temperature is, thus restricted to a much smaller range of
$T_{eff} <$ 3600 K (spectral class later than M2) and the stellar 
contribution of these stars to the
total continuum is then about 30\% or less. The line emission is extended
with respect to the neighboring continuum
making the correct identification highly probable. The OH line discussed by
Meyer et al. (1998) at 1.689 $\mu$m (red-shifted wavelength 1.696 $\mu$m) is not
obvious. This is probably due to the fact that the atmospheric
correction for that part of the 3D spectrum was not complete. Comparison
with an emission spectrum of the earth's atmosphere shows a number of
stronger atmospheric emission lines at this wavelength (Oliva \&
Origlia 1992, Maihara et al. 1993) compared to the location of the
1.542 $\mu$m OH line (red-shifted wavelength 1.548 $\mu$m). 

\subsection{\label{xxs5}Possible star formation scenarios for NGC~3227}

\begin{figure}[t!]
\begin{center}
\includegraphics[height=15.5cm,width=10.5cm,angle=-90.]{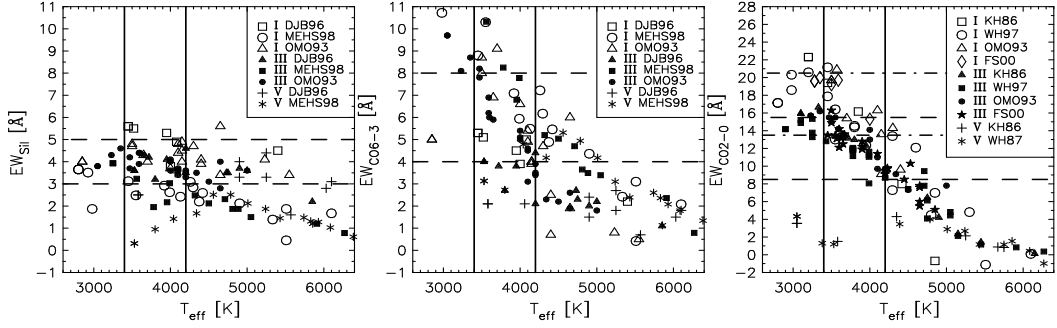}
\end{center}
\figcaption{\label{xxa11}
Determination of the luminosity class in NGC~3227. The equivalent
widths of Si I (left), CO 6-3 (middle) and CO 2-0 (right) for
different effective temperatures are shown. In each diagram the
obtained temperature range from the $\frac{CO 6-3}{Si I}$ ratio is
indicated (fat solid lines). The expected line ratios are shown by broken
lines (right figure: giants (broken line), supergiants
(dashed-dotted line)). 
(DJB96: Dallier, Boisson, Joly 1996; MEHS98: Meyer et al. 1998; 
OMO93: Origlia, Moorwood, Oliva 1993; KH86: Kleinmann \& Hall 1986;
WH97: Wallace \& Hinkle 1997; FS00: F\"orster Schreiber 2000)) 
}
\end{figure}

The equivalent width of the CO 2-0 line depends strongly on the
luminosity class of the stars -- giant or supergiant (Origlia et al.
1993). Since the slope of the nuclear spectrum suggests a large dilution
from non-stellar sources in the $H$ and $K$ band, it is not
possible to use the observed $\frac{CO 6-3}{CO 2-0}$ ratio to obtain the
luminosity class. For an effective temperature of 3600~K the expected EW for 
red giants is 14 \AA ~and 19 \AA ~for red supergiants (Fig.
\ref{xxa11}). The observed EW
of (4.0$\pm$0.9)\AA ~leads to a stellar continuum contribution from late
stars of (30$\pm$7)\% (red giant case) or (20$\pm$5)\% (red supergiant
case). 
\\
\\
Due to the dichotomy in the stellar contribution by red giants or red
supergiants to the continuum three scenarios are possible:
\\
(1) The spectrum is dominated by red supergiants. These stars peak
around an age of 15 Myr in a starburst (e.g. Schinnerer et al. 1997,
Origlia et al. 1999).
(2) AGB stars contribute significant to the NIR stellar continuum. The
largest number of AGB stars is expected for a starburst with an age of
$\sim$ 0.5 Gyr (Lancon 1999).
(3) Most of the light is coming from red giants. In this scenario an old
stellar population of $\sim$ 10 Gyr -- like the bulge population --
is present.
We investigate these three scenarios in further detail in section
\ref{xxs7} using stellar
population synthesis and NIR spectral synthesis analysis.

\section{\label{ww03}NIR IMAGING}

The high angular resolution NIR data can be used in conjunction with
optical HST data to search for structures in the circum-nuclear region
(section \ref{xxcr}). In addition we use these data to estimate the
contribution of the AGN and stellar cluster to the NIR continuum via
decomposition of the radial surface density profiles and NIR $JHK$
colors (section \ref{xxse}).

\subsection{\label{xxcr}The circumnuclear region}

\begin{figure}[t!]
\begin{center}
\includegraphics[height=15.5cm,width=10.5cm,angle=-90.]{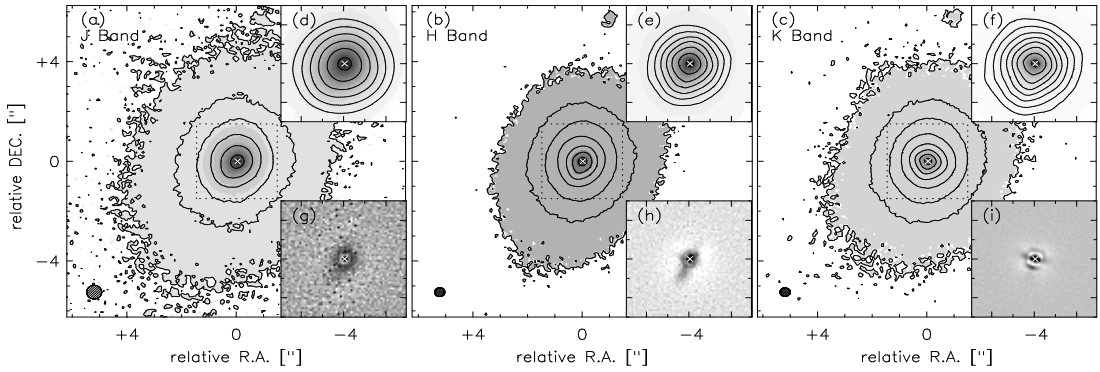}
\end{center}
\figcaption{\label{xxa01}
$J$, $H$ and $K$ band SHARP maps and smooth-subtracted maps of NGC~3227. 
For comparison the contours of the PSF reference star are overlaid on
the maps of NGC~3227 (d - f).
In the smooth-subtracted maps (g - i) no prominent structure is seen.
(a) $J$ band at 0.55'' resolution: contour lines are at 2, 4, 8, 16, 32, 
64 and 100\% of the maximum intensity of 0.08 mJy/pixel (1 Pixel = 0.05'' 
$\times$ 0.05'')
(b) $H$ band at 0.35'' resolution: contour lines are at 1, 2, 4, 8, 16, 32, 
64 and 100\% of the maximum intensity of 0.19 mJy/pixel (1 Pixel = 0.05'' 
$\times$ 0.05'')
(c) $K$ band at 0.35'' resolution: contour lines are at 0.5, 1, 2, 4, 8, 16, 
32, 64 and 100\% of the maximum intensity of 0.35 mJy/pixel (1 Pixel = 0.05'' 
$\times$ 0.05'').
The contours for the reference star are identical to the contours for
NGC~3227 in the maps with a larger FOV.
The smooth-subtracted maps of the $J$ (g), $H$ (h) and $K$ (i) band
show a 3'' FOV. 
}
\end{figure}

\begin{figure}[t!]
\begin{center}
\includegraphics[height=15.5cm,width=10.5cm,angle=-90.]{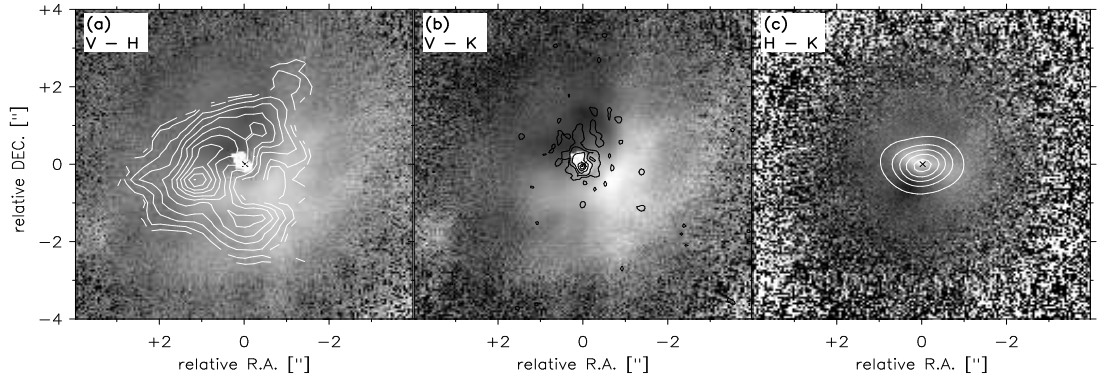}
\end{center}
\figcaption{\label{xxa05}
$V-H$, $V-K$ and $H-K$ color maps of NGC~3227 using the HST and SHARP1
data at an angular resolution of 0.35''. White in the figure corresponds to 
redder colors in the galaxy. The [O~III] line emission
from Schmitt \& Kinney (1996) is shown in the $V-K$ map in contours of
0.5, 2, 8,  32, 64 and 100\% of the maximum. The molecular line
emission of $^{12}$CO 2-1 is shown in the $V_H$ map in contours of 5,
10, 20, ... 100\% of the maximum (from Schinnerer, Eckart \& Tacconi
2000). The deconvolved \htwo 1-0 S(1) line map at a resolution of 0.8''
is shown in contours of 10, 30, ... 90 \% of the maximum in the $H-K$
map. In the $V-H$ and $V-K$
maps a red ring is visible in bright gray tone around the nucleus. In
direction of the ionization cone ([O~III] contours towards north-east)
this ring is bluer. The red east-west extension of the nucleus in the
$H-K$ map is in agreement with the extended \htwo ~line emission and
stellar absorption lines. The 0.5'' diameter ring is likely the first diffraction ring.
}
\end{figure}

A prominent dust lane is visible in the HST image running south-west
of the nucleus at a distance of about 2.5''. This structure is best
seen in the $V-H$ and $V-K$ images (Fig. \ref{xxa05}). An additional emission
knot is observed about 5'' north of the nucleus (Fig. \ref{xxa01}). The circumnuclear
region exhibits three outstanding components: a ring with $\sim$ 0.9''
radius which is highly likely related to the nuclear molecular gas ring
observed in the \CO ~line emission (Schinnerer, Eckart \& Tacconi
2000), a $\sim$ 1'' elongated structure in the east-west direction in the
$H-K$ image which is probably be due to the extended \htwo ~line emission
(Schinnerer, Eckart \& Bertoldi in prep., Quillen et al. 2000)
and a blue cone north-east of the nucleus which
coincides with the ionization cone.
\\
\\
\underline{\it The ring and the elongated structure:}
In the inner 4'' a ring segment is very prominent $\sim$ 0.9''
southwest of the nucleus. This ring-like structure surrounds the
nucleus entirely at lower magnitudes ($\sim$ 3.6$^{mag}$ (4.0$^{mag}$)
in the $V-H$ ($V-K$) map). This ring and the nucleus appear
to be connected via a $\sim$ 1'' elongated structure seen in the $H-K$ image
(Fig. \ref{xxa05}).
The ring coincides in radius and orientation with the nuclear
molecular gas ring mapped in the \COz ~line (Schinnerer, Eckart \&
Tacconi 2000). The elongated structure agrees with the orientation of
the NIR \htwo ~line emission as observed with 3D 
(Schinnerer, Eckart \& Bertoldi in prep., Quillen et al. 2000). 
\\
The highly reddened region south-west of the nucleus is likely related
to the red-shifted \CO ~emission. This region could be a molecular cloud
complex located above the plane of the galaxy (Schinnerer, Eckart \&
Tacconi 2000).
\\
We checked whether the ring could be an artifact due to misalignment of the
HST image versus our NIR data by shifting the images relative to each
other in all four directions. The ring structure is a robust 
feature even if
we allow for very large positioning errors of 0.25'' (5 pixel). 
To make sure that the elongated structure is a
real feature we made an $H-K$ color image of the reference star
where no such structure is seen. This
structure is also seen in the NIR adaptive optics data of Chapman,
Morris \& Walker (2000). 
\\
\\
\underline{\it The blue cone:}
Northeast of the nucleus the colors in the $V-H$ and $V-K$ image are
bluer by $\sim$ 1.1$^{mag}$ and $\sim$ 1.5$^{mag}$, respectively,
compared to the ones southwest of the nucleus. 
The morphology and position angle of the feature coincides with the ionization
cone mapped in its [O~III] line (Schmitt \& Kinney 1996) suggesting
that the blue cone is related to the AGN activity. The colors within
the cone are consistent with those of typical E or S0 galaxies (Frogel et al. 
1978). This could indicate that at this position we have a direct view 
onto the underlying old stellar population. Due to the high-energy AGN radiation in the cone the dust could have been sublimated or
removed to allow a direct view onto the stellar disk. The observed
difference of $\sim$ 1.1$^{mag}$ and $\sim$ 1.5$^{mag}$	in the $V-H$
and $V-K$ colors, respectively, might then
come from extincting material above the disk.
Gonzales Delgado \& Perez (1997) took optical spectra along various
slit positions across the nucleus of NGC~3227. The continuum in the
$V$ band is enhanced north-east of the nucleus as well as the H$\alpha$
and [O~III] line emission up to a radial distance of 5''. Their
analysis of the line ratios shows typical values for NLRs.
\\
\\
\underline{\it Properties of the galaxy disk:}
The NIR colors of the disk become redder with radius (see Table
\ref{xxt02}) suggesting a large
contribution of hot dust emission (up to $\sim$ 50\%) and an
extinction of $\sim$ 4$^{mag}$ compared with the colors of a typical Sc galaxy
(Frogel et at. 1978). This finding is in agreement with the
indications for higher extinction found by Barbon et al (1989). They
infer a similarly high extinction to explain the observed luminosities
of the supernova SN~1983U which occurred about 18.5'' west
of the nucleus in the disk. This high extinction value for the 
disk could be due to gas compression because of the interaction with NGC~3226 
(Mundell et al. 1995).
\\
\\
\underline{\it The northern knot:}
In all single band images an emission knot about 5.5'' north and 0.9''
west of the nucleus is apparent. In the HST image this region is
unresolved with a FWHM of $\sim$ 0.13'' (Fig. \ref{xxa01}). The NIR colors 
(Table \ref{xxt05})
lie within the region occupied by HII regions in the $JHK$ diagram 
(Glass \& Moorwood 1985). Assuming typical Sc galaxy colors there is evidence 
for extinction of $\sim$ 1.3$^{mag}$ and about 20\% dust emission
(T=600K). The measured and dereddend colors of $V-K$=2.5$^{mag}$,
$J-H$=0.6$^{mag}$ and $H-K$=0.4$^{mag}$ (translates into
$H-K$=0.2$^{mag}$ if corrected for the dust emission) are consistent
with colors expected for K stars (regardless of the luminosity
class). In the H$\alpha$ map of Gonzales Delgado \& Perez (1997) no
emission is present at this location. However, the region might lie
below their detection limit due to its small extent and
correspondingly high dilution in seeing-limited images.

\subsection{\label{xxse}The nuclear region}

\begin{figure}[t!]
\begin{center}
\includegraphics[height=15.5cm,width=10.5cm,angle=0.]{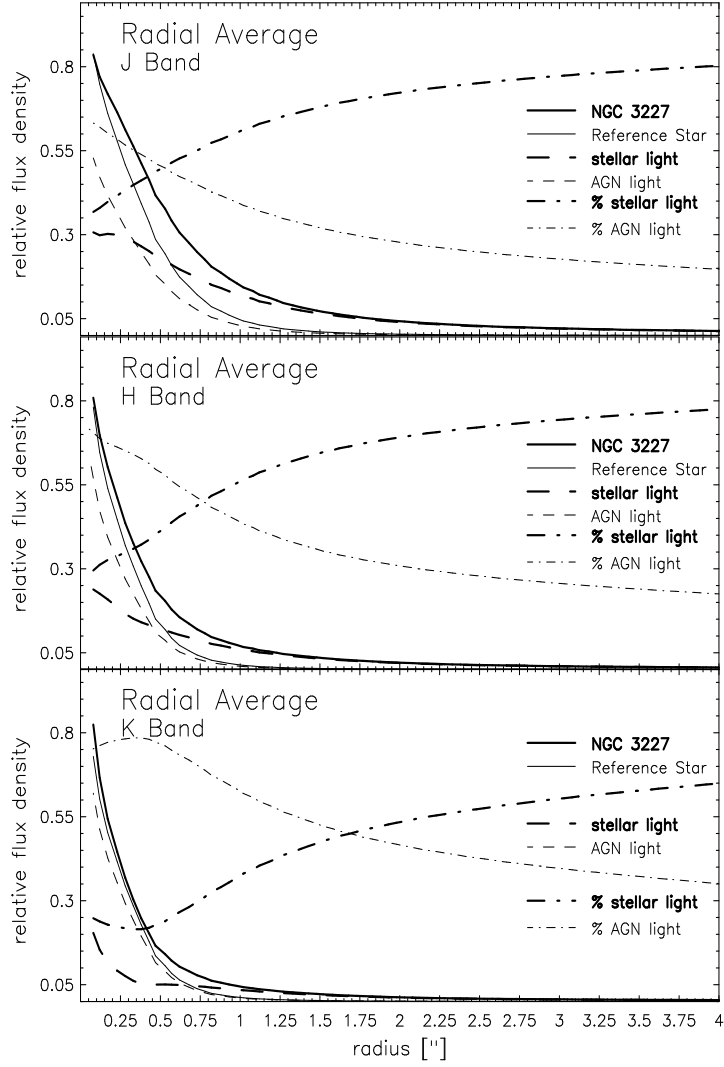}
\end{center}
\figcaption{\label{xxa02}
Comparison of the radial averages of the flux density distribution in NGC~3227 
(fat solid line) to the reference star (PSF) (thin solid line) in the NIR $JHK$ 
bands. To estimate the contribution of the nuclear stellar cluster
mapped in its absorption lines (see section \ref{xxs55}), we
subtracted the PSF representing the AGN emission from the observed
profile. The resulting stellar light profile (fat broken line) and AGN
profile (thin broken line) are shown as well. The FWHM of the stellar
light profile is $\sim$ 0.9'' if we account for the underlying bulge
light. This size is similar to the FWHM measured in the absorption
line maps (see Fig. \ref{xxa08}). In addition the relative
contribution of the stellar light (fat dashed-dotted line) and the AGN
emission (thin dashed-dotted line) to the total flux density within
each radius is shown. The radial averages were
calculated in steps of one pixel (0.05'') to obtain a good sampling
of the PSF and the galaxy nucleus. For comparison the curves were
normalized to one.
}
\end{figure}

\begin{figure}[t!]
\begin{center}
\includegraphics[height=15.5cm,width=10.5cm,angle=0.]{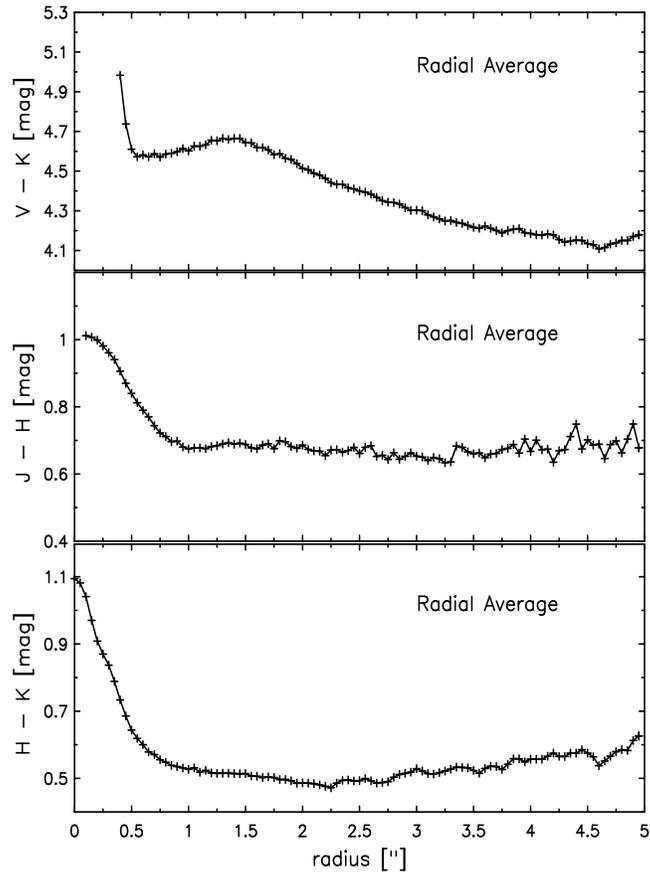}
\end{center}
\figcaption{\label{xxa03}
Comparison of the radial averages of the colors in NGC~3227. The radial
averages of the flux densities were used to obtain the colors. Due to the
saturated nucleus in the HST image, the $V-K$ color can only be derived
for radii $\geq$ 0.35''.
}
\end{figure}

\begin{figure}[t!]
\begin{center}
\includegraphics[height=15.5cm,width=10.5cm,angle=0.]{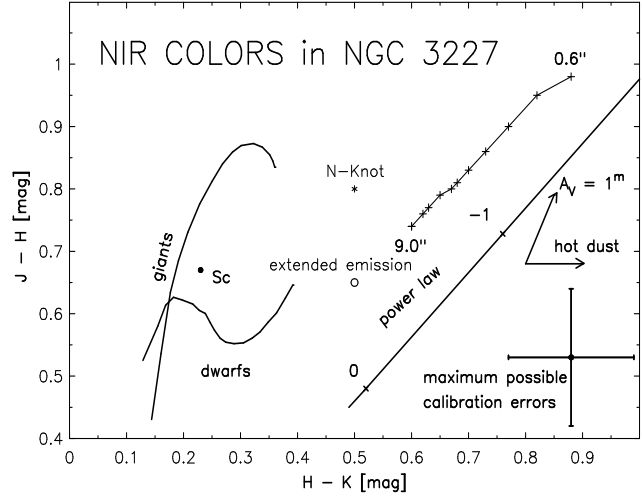}
\end{center}
\figcaption{\label{xxa04}
NIR $JHK$ diagram of NGC~3227 in full apertures of different size
(apertures of 0.6'', 1.0'', 1.4'', 1.8'', 2.2'', 2.6'', 3.0'', 3.6'',
4.6'', 6.0'', 9.0'' measured in the SHARP maps). Values of the northern
knot (star) and the extended emission ($\circ$) at a radius
of 1.0'' are shown in addition. For comparison the
typical colors of late dwarf and giant stars as well as the typical colors
of a Sc galaxy (Frogel et al. 1978) are indicated. The expected
contributions of extinction and hot dust emission as well as the power
law for pure AGN emission are shown.
}
\end{figure}

The nucleus of NGC~3227 is compact in all three NIR
bands. In the $J$ and $H$ band the nucleus is spatially resolved but not in
the $K$ band. 
This is probably due to the fact that the relative contribution
of the nuclear stellar cluster (see section \ref{xxs55}) to the nuclear NIR
continuum is increasing with shorter wavelength.
The $JHK$ colors suggest a stellar contribution of
$\sim$ (20 - 50)\% extincted by about 3 magnitudes in a 0.6'' aperture. 
The non-stellar
contribution comes very likely from hot dust emission and continuum emission
of the AGN itself. In addition an extended component is present which
can be identified with the nuclear stellar cluster seen in its stellar
absorption lines (see section \ref{xxs55}). 
Smooth-subtracted images (Fig. \ref{xxa01})
show no evidence at our achieved sensitivity of 2$\times$10$^{-5}$
Jy/arcsec$^2$ (3$\sigma$) for the nuclear spiral
structure in the inner arcsecond reported by Chapman et al. (2000).
Figure \ref{xxa01} also
shows the inner 3'' of NGC~3227 overlaid with contours of the PSF in
each band.  
\\
\\
\underline{\it The unresolved nuclear (AGN) component:}
Comparison of the radial averaged flux density profiles of NGC~3227
and the reference star reveals that most of the $J$ band source is
resolved whereas the nucleus is not resolved in the $K$ band (Fig.
\ref{xxa02}).
The contribution of the unresolved nuclear component to the overall 
flux density decreases rapidly with radius (Fig. \ref{xxa02} and \ref{xxa03}).
The very red nuclear colors of $J-H$=0.98$^{mag}$ and
$H-K$=0.88$^{mag}$ in a 0.6'' aperture decrease by about 0.5$^{mag}$
at larger radii. 
The very red colors of the nucleus (compared to the colors of a typical
Sc galaxy from Frogel et al. 1978) can be explained via two possible
scenarios: (1) A $\sim$ 50\% contribution of hot dust (T=600~K) together
with an extinction of A$_V$=3$^{mag}$ (Fig. \ref{xxa04}). (2) The AGN itself 
contributes to the $K$ band flux density with $\sim$ 50\%
(with a power law of S$_{\nu} \sim \nu^{-1.5}$) plus about 30\% from
dust emission and about 20\% stellar continuum. The second scenario
is consistent with the results of the imaging spectroscopy data
(section \ref{ww04} and \ref{xxs55}) and also the modeling results of
section \ref{xxs7}.
\\
\\
\underline{\it The resolved nuclear (stellar cluster) component:}
The radial surface brightness profiles indicate the presence of a second 
extended component contributing to the nuclear emission. This component together
with the underlying bulge is 
responsible for at least 50\% of the $JHK$ flux density
at a radius of 1'' (after subtraction of the unresolved component,
Fig. \ref{xxa02}).
Decomposing the radial profiles into a unresolved point source
representing the AGN and the resolved nuclear stellar cluster with a
FWHM of $\sim$ 0.9'' (see section \ref{xxs55}) shows that the contribution of
the nuclear stellar cluster is decreasing from $\sim$ 65\% in the $J$ band to
40\% in the $K$ band. This is consistent with the values found from
the analysis of the 3D data (section \ref{xxs55} and \ref{xxs7}). 
The FWHM for the $H$ band is in very good agreement with the FWHM of
0.9'' found for the nuclear stellar cluster (section \ref{xxs55}). The
decomposition was more difficult for the $J$ and $K$ band, since the
sensitivity of the detector in the $J$ band is lower affecting the 
proper determination of the brightest peak in the single frame maybe leading
to a broadening of the profile. The decomposition in the $K$ band was
hampered by the fact that the stellar contribution is smaller at this
wavelength and the PSF did not perfectly match the NGC~3227 data. 
The extended emission at a 1'' radius probably represents the undiluted
properties of the nuclear stellar cluster and the underlying bulge
population. Its NIR colors of $J-H$=0.65$^{mag}$ 
and $H-K$=0.50$^{mag}$ can be either explained as a $\sim$
30\% contribution due to dust emission (Fig. \ref{xxa04}) or as a young 
stellar population with an extinction of (5 - 6)$^{mag}$ 
(see curves in Fig. 8 of Hunt et al. 1997). 
\\
The NIR colors in a full 3.6'' aperture (as used in section 
\ref{xxs7} and the appendix) can be either explained by
$\sim$ 40\% dust contribution to the $K$ band plus a stellar emission
which is extincted by 1.5$^{mag}$. However, in this case no AGN
contribution to the $H$ and $J$ band is necessary, in contradiction to
the results of the decomposition of the radial surface brightness profiles
and the imaging spectroscopic data (section \ref{ww04}).
Therefore, the colors have to be explained by a $\sim$ 30\% dust
contribution plus a similar amount of AGN light together with a
younger stellar component which is highly extincted (A$_V$ $\sim$ 4-5
$^{mag}$). 

\section{\label{xxs7}AGE AND DISTRIBUTION OF THE NUCLEAR STELLAR CLUSTER}

It is important to estimate the contributions from the nuclear stellar
cluster to the total nuclear continuum emission. Winge et al. (1994) have
obtained a mean stellar contribution to the nuclear continuum 
of about $\sim$ 40\% from observations in the visible. This depends on the age 
of the stellar population and
therefore the modeling of the star formation history. 
We now look at the three scenarios mentioned in 
section \ref{xxs5} representing different ages 
using population synthesis and NIR spectral synthesis programs. A
short description of these programs together with the estimates of the
used parameter is presented in the appendix. The observed values are
summarized in Table \ref{xxt10}).
The aim is to obtain a population dominated by cool M stars consistent with the
result of the absorption line analysis (see section \ref{xxs55}). 
To scale the total continuum contribution of M stars accordingly 
additional non-stellar contributors were added in SPECSYN.

\subsection{The Model Parameters and the Fitting Procedure}

For each scenario we started by fitting the observed contribution of
cool stars to the $K$ band continuum in STARS narrowing the possible
model range. Via the stellar flux contribution
in the $K$ band, the $K$ band luminosity was adjusted for each scenario
which resulted in a SFR and enabled us to estimate the stellar mass. 
In addition the bolometric luminosity \lbol, the Lyman continuum
luminosity \llyc , and the supernova rate \nsn ~were computed for
comparison to the observed values (see appendix and Table
\ref{xxt10}).
\\
The decay time ($t_{burst}$) was set to 100 Gyr for continuous
star formation. A decay time of $t_{burst}$ = 3 Myr was used for starbursts. 
Such a time seems reasonable, as the first supernovae are expected to explode
after $\sim$ 1~Myr and to disturb the interstellar medium preventing
further star formation. Maiolino et al. (1998) investigated the effect
of different decay times on the model parameters. As seen in their
Figure 10, the difference in decay time has a strong impact on the
parameters connected to high mass stars such as Lyman continuum
luminosity or supernova rate whereas the difference in \lk/\lbol ~is
negligible (see also Figure 8b of Genzel et al. (1995) for
$t_{burst}$ = 5 Myr). Ideally the decay time $t_{burst}$ should be treated as a
free parameter, however, due to the uncertainties in the observed
quantities of \llyc ~and \nsn ~no strict constraints can be placed. 
\\
\\
In SPECSYN the fit to the spectral slope was done
simultaneously in the $H$ and $K$ band by adding extinction, dust emission and
a contribution of non-thermal AGN emission (see also section
\ref{xxnst}). 
Since SPECSYN does not include spectra of AGB stars, spectra from M giants
were used instead. For mass estimates only stars in the range of 0.80 
\solm ~($\approx$ K0 V star) to 120 \solm ~are used. 
Therefore the obtained stellar mass can only be regarded as a lower
limit assuming that the IMF is continuous as observed in many cases
(see Elmegreen 1999 for an overview).

\subsection{\label{xxnst}The Non-Stellar Contributions to the Nuclear NIR Continuum}

\begin{figure}[t!]
\begin{center}
\includegraphics[height=15.5cm,width=10.5cm,angle=-90.]{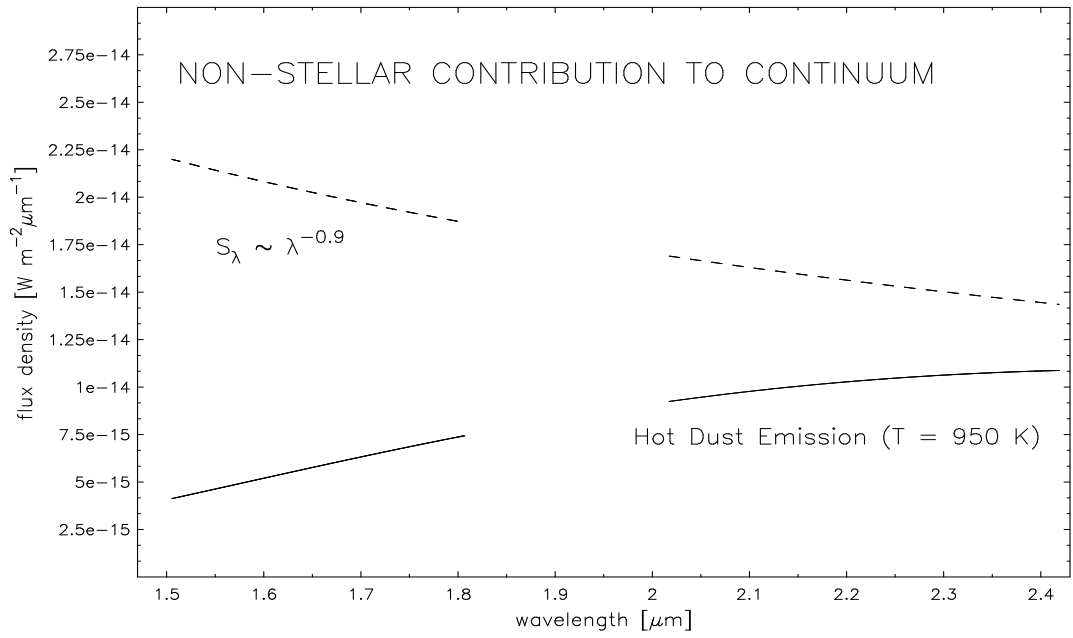}
\end{center}
\figcaption{\label{xxa13}
The non-stellar contributions to the continuum in the $H$ and $K$ band
for NGC~3227. The contribution of a 900 K hot dust emission (solid
line) and of the non-thermal AGN contribution (broken line) are
shown.
}
\end{figure}

The discrimination between the three scenarios is 
independent of the non-stellar components. The best fits to 
the spectra yielded the same percentage of non-stellar contribution in each 
case. This can be explained by the general spectral slope of cool M stars,
which dominate the stellar light, and is similar for all three scenarios. 
\\
The following procedure was adopted: (1) In the $H$ band the
AGN power law contribution to the observed spectrum was adjusted by
matching the depth of the stellar absorption features. (2) The value
of the extinction was derived by matching the slope of the observed
$H$ band spectrum. (3) The $\lambda$2.29$\mu$m CO bandheads were used
together with the $K$ band spectral slope to define the dust temperature as
well as the dust contribution. (4) The AGN power law contribution was
checked and re-adjusted to fit the observed $H$ and $K$ band spectrum.
This method is relatively robust, since the AGN power law contribution
and the extinction are derive almost independent of each other, as they
affect the $H$ band spectrum very differently. By fitting the $K$ band
spectrum the only free parameter is then the contribution of hot dust
emission which can be treated independently. Changes of $\sim$ 100~K
in the dust temperature give already no longer satisfying fits to the
spectrum despite varying the relative contribution of the dust
emission.
\\
For all three scenarios the modeling resulted in similar amounts for
the stellar contribution to the total continuum within the given
uncertainties (Fig. \ref{xxa13}). The values for the hot dust emission
($\sim$ 25\%), the AGN contribution ($\sim$ 35\%) and the extinction
(A$_V$ $\sim$ 5$^{mag}$) are in agreement with those derived from the
$JHK$ colors (see section \ref{xxse}).
In the $H$ band about 65\% and in the $K$ band about
40\% of the continuum are due to stellar light. In addition we find an
extinction of A$_V \sim$ (4-5)$^{mag}$. The difference is due
to the SEDs of the different contributing components. 
In the case of NGC~3227 this means that
the $H$ band is less affected/contaminated by emission associated with
the AGN. Although, the contribution from the power law component is larger than
in the $K$ band due to the low spectral index, the 
contribution of the hot dust emission is much larger in the $K$ band.
Changes in $\alpha$ for the AGN power law contribution have no
relevant effects on the fitting in the range of -1.1 to -0.6, since the slope 
is relative shallow.
We adopted $\alpha$ = -0.9, as this is consistent with the slope 
observed in the varying optical continuum of NGC~3227 (Winge et al. 1995).
The derived dust temperature is of the order of 950~K, in agreement
with findings from Oliva et al. (1995) who found similarly high dust
temperatures in their Seyfert sample. 
As outlined above, the derived dust temperature (plus the relative
contribution of the dust emission) is very sensitive to the longer
wavelength range of the $K$ band. Assuming, that the derived values
for the extinction and AGN power law contribution from the $H$ band
fit are correct, this leaves only a small range of about $\pm$ 70~K
for the dust temperature. However, it is very likely that two
dust emitting components are present one associated with the AGN and
the other with the stellar cluster.

\subsection{\label{xxagb}The AGB Scenario: 0.5 Gyr old Starburst}

This scenario can be discriminated from the other two using the spectra,
colors, spectral lines and the results from the population and NIR spectral
synthesis.
\\
\\
\underline{\it The spectral slope:}
Comparison with the composite spectra of Lancon (1999) shows that a 200 Myr 
stellar cluster that contains only C-rich AGB stars shows a prominent sharp 
absorption edge due to the C$_2$ $\lambda$ 1.77 $\mu$m line. Such a prominent
feature is not observed in the spectra of NGC~3227. Since the line 
is shifted at the band edge where the atmospheric calibration is poorer,
this possibility can not be completely excluded. However, the overall spectral
slope is not in agreement with the AGB spectra of Lancon (1999). 
\\
In the case of a 200~Myr old stellar cluster dominated by O-rich AGB
stars, the spectrum is dominated by deep H$_2$O absorptions between
the $H$ and $K$ band which affect the $H$ band more than the $K$ band. 
The continuum drops significantly towards the
band edges. Such a behavior is not obvious in the $H$ and 
$K$ band spectrum of NGC~3227. This indicates that O-rich AGB stars are not
the dominant spectral type. The $H$ and $K$ band spectrum obtained by 
Vanzi et al. (1998) also show no drop in flux at the band edges. Since no
synthetic spectrum showing a mixture of C- and O-rich AGB stars or showing
AGB stars at different age is given by Lan\c{c}on (1999),
this scenario can not completely be excluded from the spectral slope alone,
however. 
\\
\\
\underline{\it The colors:}
As shown by Lancon (1999) the $V-K$ and $J-K$ colors can be used to
discriminate AGB dominated populations from older (giant dominated) 
populations. 
The pure colors of the nuclear stellar cluster of NGC~3227 can be estimated
from the radial averages (Fig. \ref{xxa03}) at a radius of $\sim$ 1''
to minimize contamination from the AGN. As shown in section
\ref{xxse}, the $JHK$ colors (of the extended emission in Fig.
\ref{xxa04}) can be explained either by a young cluster with strong
extinction or 30\% dust emission in addition to an underlying bulge
component with typical Sc colors (Frogel et al. 1978). However, the underlying 
bulge component might already completely dominated the colors as suggested by 
the radial decomposition (Fig. \ref{xxa02}). The measured $J-K$ color of $\approx$ 
1.15$^{mag}$ is similar to the expected value of (0.9 - 1.0)$^{mag}$ during the 
AGB phase (Lan\c{c}on 1999). The observed $V-K$ color is
$\approx$ 4.6$^{mag}$. However, the interpretation is complicated by
the fact that
a significant fraction of the $V$ band flux density comes from the
NLR lines. This is quite obvious in the $V-K$ color map where the
ionization cone is prominent in bluer colors. To take this into
account, one has to decrease the $V$ band flux and therefore
increase the $V-K$ color. On the other hand the $V$ band flux has to
be increased due to the observed extinction (see section \ref{xxse}).
This makes it possible that the AGB phase $V-K$ colors of about 2.5$^{mag}$ 
(Lan\c{c}on 1999) can be matched. The analysis is, however, hampered
by the fact that the exact contribution of the nuclear stellar cluster
to all the individual bands is uncertain.
\\
\\
\underline{\it MgI/AlI at 2.11 $\mu$m:}
The $K$ band spectrum of NGC~3227 shows indications for absorption
lines from MgI/AlI at 2.11 $\mu$m (Fig. \ref{xxa10}). In the
corresponding line map (Fig. \ref{xxa08}) the 50\% contour of the
absorption is extended with respect to the neighboring continuum in
agreement with the other observed stellar lines. In the spectra of
three AGB stars (two C-rich, one O-rich) from Wallace \& Hinkle (1997)
no MgI/AlI absorption feature is present. The comparison of the lower
limit of the equivalent width (value minus 3$\sigma$ value) to the
expected value of an M star of 4.9 \AA ~suggests that at least about 15\% of
the total continuum can be due to M stars. Since about 40\% of the $K$
band continuum is of stellar origin, these 15\% translate then into about 40\% 
of the stellar light coming from M stars and not AGB stars. This
result is just in agreement with the models for the (peak) AGB phase where up 
to 50\% of the $K$ band continuum can be due to AGB stars (Lan\c{c}on 1999). 
However, since the EW of MgI/AlI is probably higher
the contribution from AGB stars is lower indicating that the nuclear stellar
light is not completely dominated by AGB stars.
\\
\\
\underline{Population and NIR spectral synthesis:}
Since there is an insufficient number of spectra for AGB stars available in the
literature, we approximated the AGB spectra in SPECSYN with spectra from
cool M stars (which should give a reasonable agreement in the spectral
slope). The relevant numbers of the population synthesis are given in
Table \ref{xxt10}. The percentage of the M stars to the $K$ band light is
at the lower limit of the observed value.
The present-day stellar mass is only a third of the dynamical mass in
the inner 3.6''. The estimated black hole mass is about 4$\times$10$^7$ \solm
~(Ho 1998) and the molecular gas mass is about 5$\times$10$^7$ \solm
~(Schinnerer, Eckart \& Tacconi 2000). This indicates that there must be an
additional older ($>$ 1 Gyr) stellar population which contributes
little to the observed light but a lot to the mass. To verify
our spectral synthesis we fitted the spectrum of the Mira star $\chi$
Cyg (Wallace \& Hinkle 1997) plus an additional flat continuum to the
$K$ band spectrum of NGC~3227. To obtain a reasonable fit in the
spectral slope (Fig. \ref{xxa06}) a non-stellar continuum of about 70\%
is needed in agreement with the results of the spectral synthesis.
\\
\\
In conclusion, most of the evidence does not
favor the AGB phase and therefore an age of $\sim$ 5$\times$10$^8$
yr. However, despite the remaining uncertainty in addition to the
mapped nuclear stellar cluster at least a second much older stellar
population must be present.

\subsection{The RG Scenario: Old Stellar Cluster}

\begin{figure}[t!]
\begin{center}
\includegraphics[height=15.5cm,width=10.5cm,angle=-90.]{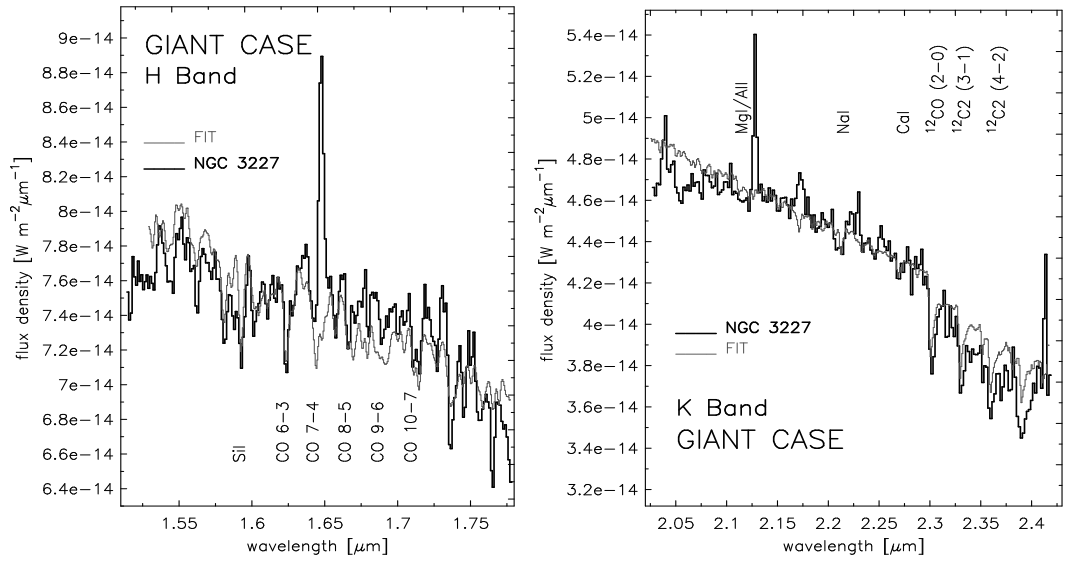}
\end{center}
\figcaption{\label{xxa14}
SPECSYN result of the RG (red giants) scenario for NGC~3227. The agreement between
the data (black line) and the synthetic spectrum (gray line) is satisfying.
}
\end{figure}

The light of an old ($\sim$ 10$^{10}$ yr) starburst would be dominated
by late type main-sequence stars. However, such a scenario would
require that no molecular gas could have recently reached the nuclear region and
be transformed into new stars. This seems unlikely given the large molecular 
gas concentrations observed close to the center (Schinnerer, Eckart \&
Tacconi 2000) and the observed nuclear activity (AGN). Therefore we assume a large age (10 Gyr) with constant star
formation which leads to a giant dominated stellar population. 
Table \ref{xxt10} summarizes the fit values.
The relative contribution from M stars in the $K$ band is smaller 
than those deduced from the stellar absorption lines. In addition to M stars, 
K stars contribute about 30\% to the stellar continuum in the $K$ band. The
resulting spectral fit (Fig. \ref{xxa14}) is not as good as in the RSG case
for the CO bandheads longword of 2.30$\mu$m. A small change in the
dust temperature can, however, minimize this effect.
The present-day stellar mass which is a lower limit due to the cut-off
in the IMF is close to the obtained
dynamical mass, if one accounts for the neglected low-mass stars. 
However, in this scenario the total observed (spatially unresolved)
Br$\gamma$ emission could be explained by the nuclear stellar cluster.
Since, the stellar absorption lines are spatially extended whereas the
Br$\gamma$ line emission is not, this indicates different origins for
the two lines. Even including extinction of the Br$\gamma$ line
flux can not solve this discrepancy in the spatial distribution. 
Therefore, continuous star formation over the last 10 Gyr
seems highly unlikely. 

\subsection{The RSG Scenario: Young $\sim$ 25 Myr Starburst}

\begin{figure}[t!]
\begin{center}
\includegraphics[height=15.5cm,width=10.5cm,angle=-90.]{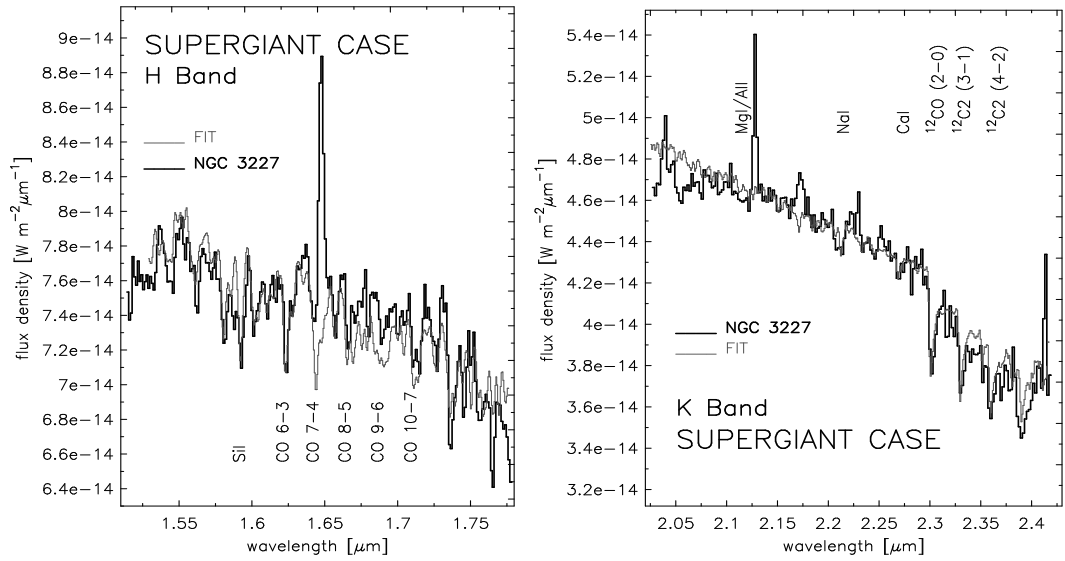}
\end{center}
\figcaption{\label{xxa15}
SPECSYN result of the RSG (red supergiants) scenario for NGC~3227. The agreement between
the data (black line) and the synthetic spectrum (gray line) is good.
}
\end{figure}

The contribution of M-type RSG stars to the stellar $K$ band light is
largest at an age of 25~Myr with about 55\% contribution. In order to
explore the age range there RSG stars dominate the NIR light, we
computed the model values at three different ages (15 Myr, 25 Myr, and
50 Myr; see Table \ref{xxt10}).
The resulting spectral fit for an age of 25 Myr is presented in Fig. 
\ref{xxa15}. 
As in the case of the AGB phase (section \ref{xxagb}) the present-day stellar mass is only a 
fraction of the total dynamical mass, again, indicating that besides that young
luminous population at least one much older and less luminous
population must be present at the center of NGC~3227 in addition to
the black hole and the molecular gas. 
The Lyman continuum luminosity \llyc ~for this burst is considerable
low for burst ages of 25 Myr and larger. Since the bolometric luminosity
\lbol ~and the supernova rate \nsn ~are similar to the observed values
for an age of 25 Myr, this indicates an age even larger than 25 Myr,
if the AGN is contributing to these quantities. However, to still
match the relative contribution of the M stars to the NIR continuum,
a burst age below 50 Myr is favored.

\subsection{Further Support for the RSG Scenario from Optical Absorption Lines}

Optical observations of the nuclear region in NGC~3227 also point 
towards a younger stellar population of RSG stars. In the optical bands
the CaII triplet at $\lambda$8498\AA, 8542\AA, 8662 \AA ~and the MgI~$b$
line at $\lambda$ 5171\AA ~can be used as a diagnostic line ratio. The
CaII lines arise in the atmospheres of late type stars and show a
similar dependency on the luminosity class as seen for the
$\lambda$2.29$\mu$m CO 2-0 line (e.g. Terlevich et al. 1990). 
Different measurements of the MgI~$b$ line towards the nucleus of
NGC~3227 indicate a stellar contribution of 17\% - 30\% (Terlevich et
al. 1990, Malkan \& Filippenko 1983, Winge et al. 1995) whereas
measurements of the CaII triplet suggest a stellar contribution of
about 60\% (Terlevich et al. 1990). This large discrepancy can be
explained if the CaII lines arise in the atmospheres of RSG stars and
not RG stars.
\\
In the spectra of Nelson \& Wittle (1995) the CaII triplet is clearly
detected whereas the MgI~$b$ line is only weak. Dilution of the MgI~$b$ line
by the neighboring emission lines of [Fe VII] and [Fe VI] can be
excluded, since the equivalent width in the high spectral resolution spectrum 
of Malkan \& Filippenko (1983) is small as well. The small
EW of the stellar absorption line of FeI $\lambda$ 5268 \AA ~which shows the 
same strength as the MgI~$b$ line in starburst galaxies 
(Malkan \& Filippenko 1983) is consistent with this finding.
\\
The varying AGN continuum can not account for 
the difference in stellar contribution derived from both lines:
The continuum varies with a period of $\sim$ 17 d (Winge et al. 1995,
Salamanca et al. 1994). However, the amplitude at 5200 \AA ~and 8500
\AA ~is only about 10\% (s. Fig. 1 of Winge et al. 1995) allowing not
for such a large difference in the stellar contribution. The
difference still persists if the CaII triplet is observed during the AGN
minimum and the MgI~$b$ line during AGN maximum. Salamanca et al.
(1994) measured the EW of MgI~$b$ in the minimum and maximum and their
value is in agreement with that of Winge et al. (1995). This confirms
that the CaII line is indeed stronger relative to the MgI~$b$, thus
suggesting a significant contribution of RSG stars to the stellar
light.
\\
Alternatively, different wavelength dependent extinction towards 
the nuclear stellar cluster could
explain the discrepancy. However, this would imply different AGN
contributions at the visible and red wavelength range in
contradiction to the observed flat slope of the variable AGN continuum
at these wavelengths (Winge et al. 1995).

\section{\label{ww08}SUMMARY AND CONCLUSIONS}

In the NIR continuum of NGC~3227 we find besides the unresolved AGN
component a stellar nuclear cluster with a FWHM of $\sim$ 70~pc.
At least two stellar populations with different ages are observed
in the nucleus of NGC~3227. Continuous star formation can be ruled
out implying that the nuclear star formation takes place in episodes.
This is in agreement with the large reservoir of molecular material at
a radial distance of $\sim$ 100~pc from the nucleus and also
associated with the HII regions in the inner 1~kpc. Since the gas
dynamics in the inner 1~kpc are complex (Schinnerer, Eckart
\& Tacconi 2000), the probability that gas will become instable and
form stars is quite high. 
We find more evidence for a nuclear stellar cluster 
with an age of 25 to 50~Myr when most of the NIR stellar light comes from 
RSG stars rather than an age of $\sim$ 5$\times$10$^8$yr where the NIR
light is dominated by AGB stars.

\begin{itemize}

\item
The nuclear stellar light in the $H$ and $K$ band is dominated by
emission from very cool evolved stars. The analysis of the stellar
absorption lines indicates M giants or supergiants as the dominant
stellar type. From our population and spectral synthesis we can rule
out constant star formation over the past 10 Gyr which could produce
an M giant dominated population. 

\item
Investigation of the other two likely scenarios (RSG phase with an age
of $\sim$ 25~Myr or AGB phase with an age of 5$\times$10$^8$yr)
shows that none can be ruled out completely. However, supporting evidence 
for the RSG scenario is found from optical observations which point towards a
large contribution of RSG stars to the stellar light. In the RSG
scenario the nuclear stellar cluster produces most of the MIR/FIR emission,
a substantial fraction of the H recombination line fluxes and the
radio flux. This implies that a considerable amount of the nuclear
luminosity is coming from the starburst and not the AGN component.

\item
Although the AGB scenario can not be excluded, the observed parameters
are at the uncertainty limits for this model. Due to the
different spatial resolution of the imaging and imaging spectroscopic
data a more detailed analysis is not possible. In addition it seems
unlikely that no nuclear star formation has occurred in the last 10$^8$
yr given the large molecular gas reservoir close to the inner 3.6''.
To evaluate this further data at higher S/N and with a larger coverage
in the wavelength range are needed. 

\item
It is interesting to note that in the case of either the RSG or AGB
scenario the mapped nuclear stellar cluster accounts for only $\sim$
15\% of the dynamical mass in the inner 3.6''. As the mass
contribution of the black hole and the molecular gas is of the same
order, this implies that a
least another stellar population exists which is much older and less
luminous. Such a scenario is similar to the center of our own galaxy
where stars from several star formating episodes are observed (Krabbe
et al. 1995). This is also consistent with the findings of Tecza et
al. (2000) for NGC~6240. In a possible scenario the star formation and
AGN activity might go through cycles of high and low activity
depending on whether the nuclear starburst is providing fuel for the AGN or
not.

\item
The stellar contribution to the continuum emission is largest in the
$H$ band with 65\%, whereas the fraction of the stellar light is only
$\sim$ 40\% in the optical and $K$ band. The results of the spectral
synthesis are in agreement with the findings of the decomposition of the 
NIR radial flux density profiles.
The extended component shows a significant extinction of
A$_V$ $\sim$ (4 - 5)$^{mag}$ assuming young stars (Hunt et
al. 1997). The nuclear AGN component (0.5'') suggests an extinction of
$\sim$ 3$^{mag}$ as well. However, the analysis of the $JHK$ colors is
strongly affected by the co-existence of an old not luminous and a young stellar
population and the AGN contributions in the inner 500~pc.

\item
In our high angular resolution NIR data the nucleus is slightly
resolved at $J$ and $H$ band, clearly showing the nuclear stellar
cluster as an additional extended component. $VJHK$ color images show the
structure of the ionization cone as well as enhanced extinction
south-west of the nucleus.

\end{itemize}

\acknowledgements
We would like to thank the MPE 3D team (N. Thatte, L.J. Tacconi-Garman, H.
Kroker, S. Anders, R. Maiolino, J. Gallimore)
for obtaining the data and the staff of
the WHT and the Calar Alto 3.5~m for their support. The NTT staff was
very helpful during the SHARP observations. 
For fruitful discussions we like to thank R. Genzel. L. Armus'
comments helped to improve and clarify the manuscript.
We thank H. Schmitt for 
kindly providing us with the HST [O~III] map of NGC~3227.

\appendix

\large
\begin{center}
{\bf APPENDIX}
\end{center}
\normalsize

\section{\label{ww06}STELLAR POPULATION AND NIR SPECTRAL SYNTHESIS}

Here we give a short description of the stellar population
synthesis code STARS together with a discussion of general caveats. 
We also outline the structure of the NIR spectral synthesis code SPECSYN.
The observational parameters which are used for modeling are derived 
in this section as well.

\subsection{Population synthesis}

The aim of the population synthesis is to obtain a model reflecting
the star formation history of the galaxy. Theoretical stellar
evolutionary tracks are used in conjunction with an initial stellar
mass function (IMF). It is assumed that stars with M $>$ 8 \solm
~explode as supernovae after their supergiant phase. Some also
experience the Wolf-Rayet phase during which strong stellar winds
influence the evolution of the star. Stars with masses between 8 \solm
~and 30 \solm ~become red supergiants which have
strong contributions in the NIR due to their low effective
temperatures. Stars with masses below 8 \solm ~evolve into red giants
where stars with masses between 2 \solm ~ and 7 \solm
~reach the asymptotic giant branch. At this point they show a strong
luminosity increase at very low effective temperatures (AGB phase), and then
evolve to planetary nebulae.
\\
\\
\underline{\it High-mass stars:}
The evolutionary tracks for high-mass stars have uncertainties at
different evolutionary stages regarding the time-span and the related physical parameters
such as effective temperatures, luminosity and mass loss. The rotation
of massive stars can, e.g., reduce their mass loss rate and therefore
change their lifetime on the main sequence (Langer 1998). Talon et al.
(1997) have investigated the influence of the rotation velocity for a 12
\solm ~star. At the same effective temperature the star showed a
higher luminosity at high rotation velocities (of 300 km/s). Therefore
knowledge of the mean rotation velocity is necessary to calculate the
stellar evolutionary tracks, since the mass loss during the main
sequence influences the evolution of the star at later
stages. The evolutionary tracks for stars with 15
\solm~$<$ M $<$ 30 \solm ~never reach low effective temperatures of T
$\sim$ 3000~K during the red supergiant phase, although such stars are
observed (Langer \& Maeder 1995). 
However, effective temperatures of about 3500~K equivalent of early M 
supergiants are reached. The depths of the prominent NIR absorption lines are
relatively similar for M supergiants as well as the NIR spectral slope
compared to the hotter stars. Therefore the effects on the NIR spectral
synthesis are sufficiently small. Only the difference in luminosity
can affect the population synthesis by underestimating both bolometric
and $K$ band luminosity. However, again the scatter within the M
supergiants is reasonably small, if the ratio of \lk/\lbol ~is
considered for M supergiants (ranging from 0.11 (M0I) to 0.15 (M5I)).
As discussed in section \ref{xxs55} the metallicity derived for the
nuclear stellar cluster in NGC~3227 is close to solar, thus justifying
to use the evolutionary tracks for solar metallicity which give
reasonable results for the red supergiant phase (Origlia \& Oliva 2000).
Understanding the evolution during the Wolf-Rayet
phase is also in progress, and models include effects such as line
blanketing (e.g. adding atomic lines like O, Gr\"afener et al. 1998)
and clumping in the stellar winds (Nugis et al. 1998). These effects
influence the mass loss and therefore the lifetime of a Wolf-Rayet
star. 
Since the ages relevant for our analysis are considerably longer
than the lifetimes of WR stars, this effect can be neglected.
\\
\\
\underline{\it Binary stars:}
Van Bever \& Vanbeveren (1998) have investigated the effect of binary stars
on the evolution of star forming regions via number population
synthesis calculations. The mass transfer in close binaries, so called
'accretion stars', can make these stars appear in the stellar
population as O type stars at a time at which normal O type stars would
have already
evolved away from the main sequence. This is a natural explanation
for the 'blue stragglers' found in a number of stellar clusters. In
this case the star forming region is older, as suggested
by the appearance of the accretion stars. However, the number of
accretion stars depends strongly on the abundance of binary stars and
the mass transfer via the Lagrange point L$_1$ with 'Roche lobe
overflow'. Since the correct handling of the binary star evolution is not
clear, this effect is not taken into account by STARS.
\\
\\
\underline{\it Low-mass stars:}
In the evolutionary tracks of low-mass stars the less understood phase
is the asymptotic giant branch phase (AGB-phase). AGB stars appear at a
similar location as red giants in the HR-diagram, and the
progenitor stars are more massive. This means that these stars reach
their location in the HRD much earlier. Modeling of the AGB phase at
the tip of the giant branch just before the transition to a planetary
nebula is difficult, since the stars change their luminosity,
effective temperature and chemical composition. NIR observations show
that AGB stars can change their effective temperatures by 500~K and their 
luminosities by 25\% within short time-scales (Lancon 1999). Spectra of AGB
stars are characterized by deep molecular absorption bands of H$_2$O (O-rich
AGB stars) and C$_2$ (C-rich AGB stars) (Lancon 1999 and references
therein).
\\
Lancon (1999) investigates the influence of AGB stars on the observable
parameters of stellar populations using the population synthesis code
PEGASE (Fioc \& Rocca-Volmerange 1997). She compares a synthetic
spectrum of a 4~Gyr old giant-dominated population to a 200 Myr old
(C- or O-rich) AGB stars dominated population. The continuum clearly
shows the strong molecular absorption features of the AGB stars in the
later case. About 0.3 -1.5 Gyr after an intense burst of star
formation the contribution of AGB stars to the total emission of the
stellar population is the largest (Lancon 1999), and can be as high as
about 50\% to the total $K$ band light. 
\\
The stellar evolutionary tracks used by STARS have been extended to
the TP-AGB phase using the models of Bedijn (1988) similar to the method of 
Lan\c{c}on \& Rocca-Volmerange (1996). A detailed
description is given by Schreiber (1998).
\\
\\
\underline{\it The initial mass function (IMF):}
The initial mass function (IMF) describes the number of stars formed
per mass interval. For our calculations with STARS we have
used a power law ($\sim$ M$^{-\alpha}$)
with a typical value of $\alpha$ = -2.35 for stars between about 0.80
\solm ~and 120 \solm ~(Salpeter 1955, Leitherer 1996, review by Elmegreen 
1999). For lower masses the IMF shows a flatter slope (Scalo 1986).
The analysis is complicated by the fact that the emission from low-mass stars 
is weak in the NIR and therefore harder to analyze. 

\subsection{The NIR spectral synthesis}

\begin{figure}[t!]
\begin{center}
\includegraphics[height=15.5cm,width=10.5cm,angle=0.]{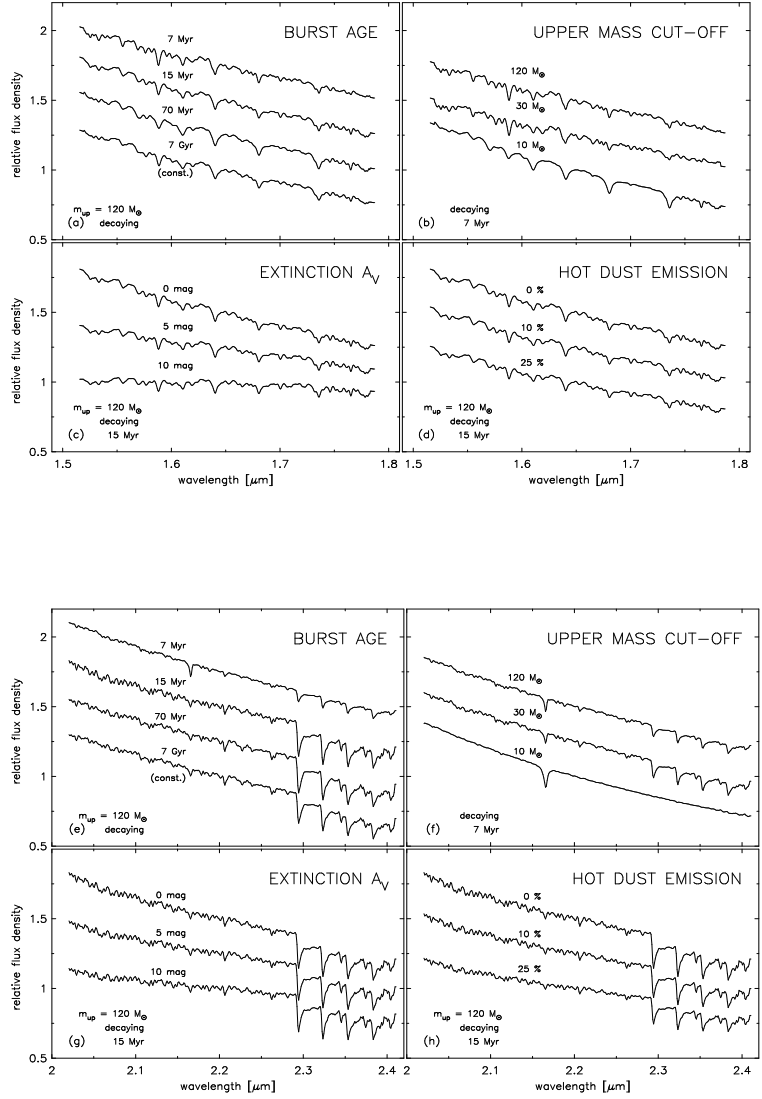}
\end{center}
\figcaption{\label{xxa12}
Influence of the different model parameter for the $H$ band (a - d)
and $K$ band (e - h) synthesized spectra of stellar clusters. All
spectra are normalized to one at 1.65 $\mu$m and 2.2 $\mu$m,
respectively, and moved by 0.25 for display. An overview of the varied
model parameter is presented in Table \ref{xxt09}.
}
\end{figure}

We test the results from our population synthesis analysis with
the NIR spectral synthesis program SPECSYN by including the contribution 
from non-stellar continuum sources to the NIR light.
SPECSYN uses the HR-diagrams to synthesize spectra by combining standard stellar
spectra from the literature at the instrumental resolution of 3D.
A detailed description of SPECSYN is given in the appendix of Schinnerer
et al. (1997).  
\\
For the purpose of this paper SPECSYN has been expanded to also
synthesize $H$ band spectra. We have implemented the possibility to add AGN
emission as a power law. 
The coverage in the $H$ band is achieved with the stellar library from Meyer
et al. (1998) and in the $K$ band with the library from Wallace \& Hinkle (1997).
The advantage is that all spectra have the same wavelength coverage
and were obtained with the same instrument representing a homogeneous data
set. The HR-diagram coverage is done in analogy to Schinnerer et al. (1997).
The influences of the various parameters on the $H$ and $K$
band spectra are shown in Fig. \ref{xxa12}, and the parameters 
are summarized in Table \ref{xxt09}.
\\
\\
\underline{\it The non-stellar contributors:}
Extinction, dust emission or AGN continuum emission can be a possible source of
dilution to the stellar continuum. The non-stellar
contributors are included as percentages to the total flux at the central
wavelength in the $H$ and/or $K$ band. In the case of simultaneous spectral 
synthesis the $K$
band percentage values are given and interpolated to values for the $H$ band.
Extinction and dust emission are treated as given by Schinnerer et
al. (1997). The AGN is not only heating the surrounding dust 
(leading to thermal emission in the MIR/FIR) but also has a strong continuum in the
UV and optical range. 
For the AGN continuum emission, we assume a power law of 
the form $I_{AGN} (\lambda) \sim \lambda^{\alpha}$, where $\alpha$ is a
free parameter with typical values of -0.8 for the wavelength range
relevant here. Malkan \& Filippenko (1983) and Edelson \& Malkan
(1986) find values for $\alpha$ between -0.9 and -0.64 by fitting the
non-stellar continuum of a sample of Seyfert galaxies. As mentioned by the 
authors this slope is very similar to quasars.
However, it is hard to
isolate the pure AGN contribution, since it consist of at least two emission
components: The Balmer continuum and the UV black body (T $>$ 26000K)
(Edelson \& Malkan 1986). Sanders et al. (1989) speculate that the
different components might arise at different distances from the
central engine: the UV is dominated by light from the accretion disk,
whereas the IR is dominated by light from the outer parts of the disk
at radii of 0.1~pc to 1~kpc and is emitted by dust in the host galaxy.

\subsection{\label{xxop}The observational parameters}

The {\it bolometric luminosity} 
is assumed to be similar to the IR luminosity, since most of the 
UV energy is transformed to IR wavelengths
due to the photo-electric heat mechanism (Tielens \& Hollenbach
1985). The IR luminosity can be derived using the IRAS fluxes and the
equation given by Sanders \& Mirabel (1996). From the IRAS fluxes at
12$\mu$m (0.62 Jy), 25 $\mu$m (1.75 Jy), 60 $\mu$m (7.84 Jy) and 100
$\mu$m (16.93 Jy) and using 

$L_{bol}[L_{\odot}] = 5.617 \times 10^5 \times D[Mpc]^2 \times (13.48
f_{12\mu m}[Jy] + 5.16 f_{25 \mu m}[Jy] + 2.56 f_{60 \mu m}[Jy] +
f_{100 \mu m}[Jy])$

\noindent
the observed bolometric
luminosity is $\sim$ 9.3$\times$10$^9$ \solar. The 10 $\mu$m map of 
Bushouse, Telesco \& Werner (1998) shows an unresolved source in the
inner 400~pc of NGC~3227. Their 100 $\mu$m map shows that the total
emission is associated with NGC~3227. This suggests that most, if not
all, of the IRAS fluxes are coming from the inner 400~pc and this is a
good approximation for the nuclear bolometric luminosity, neglecting
high-energy AGN contributions. The estimated derived bolometric
luminosity can, however, still have AGN contributions and is therefore
regarded as an upper limit.
\\
The {\it $K$ band luminosity} comes from cool and therefore evolved
stars. We use the definition 

$L_K[L_{\odot}] = 1.16 \times 10^4 \times D[Mpc]^2 \times S_K[mJy]$

\noindent
given by Genzel et al. (1995) with a $K$
band width of 0.6 $\mu$m. 
\\
The {\it Lyman continuum luminosity} is associated with the HII regions
around hot stars, giving a measure for the contribution from young, hot
stars. Using the observed Br$\gamma$ line flux the Lyman continuum
luminosity can be estimated via

$L_{Lyc}[L_{\odot}] = 5.37 \times 10^{19} \times F_{Br\gamma}[erg
s^{-1} cm^{-2}]\times D[Mpc]^2$

As discussed in section \ref{xxs4} most of the nuclear Br$\gamma$ line
emission in NGC~3227 is associated with the BLR and NLR. Only a small
percentage of the Br$\gamma$ line flux comes from HII regions around hot stars.
\\
To estimate the {\it supernova rate} we use the empirical relation
given by Condon (1992). Huang et al (1994) found a similar number by
direct comparison of the SN numbers to the radio emission in the
starburst galaxy M~82. To obtain $\nu_{SN}$ we 
used the total nuclear 5~GHz flux density of Mundell et al. (1995b) and 

$\nu_{SN}[yr^{-1}] = 3.3 \times 10^{-6} \times S_{5 GHz}[mJy] \times
D[Mpc]^2$

Since the 5~GHz flux density of Mundell et al. (1995b) might have contributions
from the AGN, the derived supernova rate is an upper limit.
All values are summarized in Table \ref{xxt10}.

\clearpage

\begin{center}
{\Large \bf References}
\end{center}

\rf{Arribas, S., Mediavilla, E., 1994, \apjj {437}{149}}
\rf{Barbon, R., Ciatti, F., Iijima, T., Rosino, L., 1989, \asa {214}{131}}
\rf{Barvainis, R., 1990, \apjj{353}{419}}
\rf{Bedijn, P.J., 1998, \asa {205}{105}}
\rf{Bushouse, H.A., Telesco, C.M., Werner, M.W., 1998, \ajj {115}{938}}
\rf{Calzetti, D., 1997, \ajj {113}{162}}
\rf{Christou, J.C., 1991, Exp. Astro., 2, 27}
\rf{Colina, L., P\'{e}rez-Olea, D.E., 1995, \mn {277}{845}}
\rf{Condon, J. J., 1992, \arasa{30}{575}}
\rf{Dallier, R., Boisson, C., Joly, M., 1996, \asas{116}{239}}
\rf{De Robertis, M.M., Hayhoe, K., Yee, H.K.C., 1998, \apjjs {115}{163}}
\rf{De Vaucouleurs, G, de Vaucouleurs, A., Corwin, H., Buta, R.,
    Paturel, G., Fougu\'{e}, P., 1991, Third Reference Catalogue of
    Bright Galaxies, Berlin, Springer-Verlag)}
\rf{Devereux, N.A., 1989, \apjj {346}{126}}
\rf{Eckart, A., van der Werf, P.P., Hofmann, R., Harris, A.I., 1994,
    \apjj{424}{627}}
\rf{Edelson, R.A., Malkan, M.A., 1986, \apjj{308}{59}}
\rf{Elmegreen, B.G., 1999, 'Unsolved Problems in Stellar Evolution',
    ed. M. Livio, Cambridge Univ. Press, in press}
\rf{Fioc, M., Rocca-Volmerange, B., 1997, \asa {326}{950}}
\rf{Fischer, J., Geballe, T.R., Smith, H.A., Simon, M., Storey,
    J.W.V., 1987, \apjj {320}{667}}
\rf{F\"orster Schreiber, N.M., 2000, astro-ph/0007324}
\rf{Forbes, D.A., Ward, M.J., 1993, \apjj {416}{150}}
\rf{Frogel, J.A., Persson, S.E., Aaronson, M., Matthews, K., 1978,\apjj{220}{75}}
\rf{Garcia, A.M., 1993, \asas {100}{47}}
\rf{Genzel, R., Weitzel, L., Tacconi-Garman, Blietz, M., Krabbe, A.,
    Lutz, D., Sternberg, A., 1995, \apjj{444}{129}}
\rf{Glass, I.S., Moorwood, A.F.M., 1985, \mn{214}{429}}
\rf{Gonz\'{a}lez Delgado, R.M., Perez, E., 1997, \mn {284}{931}}
\rf{Gr\"afener, G., Hamann, W.-R., Hillier, D.J., Koersterke, L.,
    1998, \asa {329}{190}}
\rf{Ho, L.C., 1998, Invited review to ''Observational Evidence for
    Black Holes in the Universe'', ed. S.K. Chakrabarti (Dordrecht:
    Kluwer), in press)}
\rf{Hofmann, R., Brandl, B., Eckart, A., Eisenhauer, F., Tacconi-Garman, L.E.,
    1995, SPIE-Conference, Orlando}
\rf{Huang, Z.P., Thuan, T.X., Chevalier, R.A., Condon, J.J., Yin,
    Q.F., 1994, \apjj {424}{114}}
\rf{Hunt, L.K., Malkan, M.A., Salvati, M., Mandolesi, N., Palazzi, E.,
    Wade, R., 1997, \apjjs{108}{229}}
\rf{Kleinmann, S.G., Hall, D.N.B., 1986, \apjjs{62}{501}}
\rf{Kotilainen, J.K., Ward, M.J., Boisson, C., De Poy, D.L., Bryant,
    L.R., Smith, M.G., 1992, \mn{256}{125}}
\rf{Krabbe, A., Sternberg, A., and Genzel, R., 1994, \apjj{425}{72}}
\rf{Lan\c{c}on, A., 1999, 'Asymptotic Giant Branch Stars', IAU
    Symposium 191, in press, astro-ph9810474}
\rf{Langer, N., 1998, \asa {329}{551}}
\rf{Langer, N., Maeder, A., 1995, \asa {295}{685}}
\rf{Leitherer, C., 1996, in ASP.conf. series Vol.98, p. 373.}
\rf{Lord, S., 1992, NASA Technical Memorandum 103957, Ames Research
    Center, Moffett Field, CA}
\rf{Lumsden, S.L., Puxley, P.J., 1995, \mn {276}{723}}
\rf{Maihara, T., Iwamuro, F., Yamashita, T., Hall, D.N.B., Cowie,
    L.L., Tokunaga, A.T., Pickles, A., 1993, \pasp {105}{940}}
\rf{Maiolino, R., Krabbe, A., Thatte, N., Genzel, R., 1998, \apjj {493}{650}}
\rf{Malkan, M.A., 1988, Adv. Space Res., Vol. 8, No. 2-3, 49}
\rf{Malkan, M.A., Filippenko, A.V., 1983, \apjj {275}{477}}
\rf{Malkan, M.A., Gorjian, V., Tam, R., 1998, \apjjs {117}{25}}
\rf{McAlary, C.W., McLaren, R.A., Gonegal, R.J., 1983, \apjjs {52}{341}}
\rf{Meyer, M.R., Edwards, S., Hinkle, K.H., Strom, S.E., 1998,
    \apjj{508}{397}}
\rf{Mouri, H., Nishida, M., Taniguuchi, Y., Kawara, K., 1990, \apjj {360}{55}}
\rf{Mulchaey, J.S.,Regan, M.W., Kundu, A., 1997, \apjjs {110}{229}}
\rf{Mundell, C.G., Holloway, A.J., Pedlar, A., Meaburn, J., Kukula, M.J.,
    Axon, D.J., 1995a, \mn {275}{67}}
\rf{Mundell, C.G., Pedlar, A., Axon, D.J., Meaburn, J., Unger, S.W., 1995b,
    \mn {277}{641}}
\rf{Nelson, C.H., Whittle, M., 1995, \apjjs {99}{67}}
\rf{Norman, C., Scoville, N., 1988, \apjj{332}{124}}
\rf{Nugis, T., Crowther, P.A., Willis, A.J., 1998, \asa {333}{956}}
\rf{Oliva, E., Moorwood, A.F.M., 1990, \apjjl {348}{L5}}
\rf{Oliva, E., Moorwood, A.F.M., Danziger, I.J., 1989, \asa {214}{307}}
\rf{Oliva, E., Moorwood, A.F.M., Danziger, I.J., 1990, \asa {240}{453}}
\rf{Oliva, E., Origlia, L., 1992, \asa {254}{466}}
\rf{Oliva, E., Origlia, L., 1998, \asa {332}{46}}
\rf{Oliva, E., Origlia, L., Kotilainen, J.K., Moorwood, A.F.M., 1995,
    \asa {301}{55}}
\rf{Origlia, L., Ferraro, F.R., Fusi Pecci, F., Oliva, E., 1997, 
    \asa {321}{859}}
\rf{Origlia, L., Goldader, J.D., Leitherer, C., Schaerer, D., Oliva, E.,
    1999, apjj{514}{96}}
\rf{Origlia, L., Moorwood, A.F.M., Oliva, E., 1993, \asa {280}{536}}
\rf{Ohsuga, K., Umemura, M., 1999, \apjjl {521}{13}}
\rf{P\'{e}rez-Olea, D.E., Colina, L., 1995, \mn {277}{857}}
\rf{Regan, M.W., Mulchaey, J.S., 1999, \ajj {117}{2676}}
\rf{Rubin, V.C., Ford, W.K. 1968, \apjj{154}{431}}
\rf{Salamanca, I. et al., 1994, \asa {282}{742}}
\rf{Salpeter, E. E., 1955, \apjj{121}{161}}
\rf{Sanders, D.B., Mirabel, I.F., 1996, \arasa{34}{749}}
\rf{Sanders, D.B., Phinney, E.S., Neugebauer, G., Soifer, B.T.,
    Mattews, K., 1989, \apjj {347}{29}}
\rf{Scalo, J.M., 1986, Fundamentals of Cosmic Physics, vol. 11, 1}
\rf{Scalo, J.M., 1999, 'The Birth of Galaxies', Blois, France, in
    press, astro-ph/9811341}
\rf{Schinnerer, E., Eckart, A., Quirrenbach, A., B\"oker, T.,
    Tacconi-Garman, L. E., Krabbe, A., Sternberg, A., 1997, \apjj{488}{174}}
\rf{Schinnerer, E., Eckart, A., Tacconi, L.J., 1999, \apjjl {524}{L5}}
\rf{Schinnerer, E., Eckart, A., Tacconi, L.J., 2000, \apjj {533} {826}}
\rf{Schmitt, H.R., Kinney, A.L., 1996, \apjj{463}{498}}
\rf{Schmitt, H.R., Kinney, A.L., Calzetti, D., Storchi Bergmann, T.,
    1997, \ajj {114}{592}}
\rf{Schreiber, N.M., 1998, Ph.D. thesis, LMU M\"unchen}
\rf{Shlosman, I., Frank, J., Begelman, M.C., 1989, Nature, 338, 45}
\rf{Simpson, C., Forbes, D.A., Baker, A.C., Ward, M.J., 1996, \mn
    {283}{777}}
\rf{Stasi\'{n}ska, G., Leitherer, C., 1996, \apjjs {107}{734}}
\rf{Talon, S., Zahn, J.-P., Maeder, A., Meynet, G., 1997, \asa {322}{209}}
\rf{Tecza, M., Genzel, R., Tacconi, L.J., Anders, S., Tacconi-Garman,
    L.E., Thatte, N., 2000, \apjj{submitted}{}}
\rf{Terlevich, E., D\'{i}az, A.I., Terlevich, R., 1990, \mn {242}{271}}
\rf{Thatte, N.A., Kroker, H., Weitzel, L., Tacconi-Garman, L.E., Tecza, M.,
    Krabbe, A., Genzel, R., 1995, Proc. SPIE Vol. 2475, 228}
\rf{Thatte, N., Quirrenbach, A., Genzel, R., Maiolino, R., Tecza, M.,
    1997, \apjj {490}{238}}
\rf{Tielens, A.G.G.M., Hollenbach, D.J., 1985, \apjj{291}{747}}
\rf{Van Bever, J., Vanbeveren, D., 1998, \asa {334}{21}}
\rf{Vanbeveren, D., De Donder, E., Van Bever, J., Van Rensbergen, W.,
    De Loore, 1998, New Astr., 3, 443}
\rf{Vanzi, L., Alonso-Herrero, A., Rieke, G.H., 1998, \apjj {504}{93}}
\rf{Wallace, L., Hinkle, K., 1997, \apjjs {111}{445}}
\rf{Ward, M.J., Elvis, M., Fabbiano, G., et al., 1987, \apjj {315}{74}}
\rf{Weitzel, L., Krabbe, A., Kroker, H., Thatte, N., Tacconi-Garman, L. E.,
    Cameron, M., Genzel, R., 1996, \asas {119}{531}}
\rf{Wilson, A.S., 1997, in ''Emission Linies in Active Galaxies: New
   Methods and Techniques'', ASP Conf. Series, Vol. 113, 264}
\rf{Winge, C., Peterson, B.M., Horne, K., Pogge, R.W., Pastoriza, M.G.,
    Storchi-Bergmann, T., 1995, \apjj {445}{680}}

\clearpage

\begin{table}[htb]
\caption{
\label{xxt01}
Log of the NIR observations of NGC~3227}
\begin{center}
\begin{tabular}{llrrrl}\hline \hline
Camera &             & $J$ & $H$ & $K$ & Observing run \\ \hline
3D      & T$_{int}$  &         & 1840 s & 3200 s & 3.5~m CA, Dec. 1995\\
        & T$_{int}$  &         & 1160 s & 3910 s & WHT, Dec./Jan. 1995/96 \\
	& Seeing$^{2}$ &         & $\sim$ 1.6''  & $\sim$ 1.3''  & both runs \\ \hline
SHARP 1 & T$_{int}$  &  3750 s & 1900 s & 4200 s & NTT, June 1996\\ 
        & Seeing$^{1}$ &  0.55'' & 0.35'' & 0.35'' & NTT, June 1996 \\ 
\hline \hline
\end{tabular}
\end{center}
$^{1}$ derived from the PSF (reference star)
\\
$^{2}$ derived from comparison of reference star and source data to
       SHARP~1 data
\\
The integration times are on-source observing time. 
(NTT = New Technology Telescope, La Silla, Chile; 3.5~m CA = 3.5~m Telescope
Calar Alto, Spain; WHT = William Herschel Telescope, La Palma, Spain)
\end{table}

\begin{table}[htb]
\caption{
\label{xxt06}
Data from the literature for the 3D data of NGC~3227}
\begin{center}
\begin{tabular}{lrrrl}\hline \hline
Line & Aperture & Flux$_{3D}$$^c$ & Flux$_{lit}$ &Reference \\ 
      &         & [10$^{-17}$ W m$^{-2}$] & [10$^{-17}$ W m$^{-2}$] & \\ \hline
H$_2$ 1-0 S(1) & 5.4'' & 3.02 $\pm$ 0.17 & 3.3 $\pm$ 0.6 & Fischer et al.  1987 \\
H$_2$ 1-0 S(0) & 5.4'' & 0.52 $\pm$ 0.11 & 2.1 $\pm$ 0.5 & Fischer et al.  1987 \\
H$_2$ 2-1 S(1) & 5.4'' & 0.49 $\pm$ 0.15 & 2.6 $\pm$ 1.3 & Fischer et al.  1987 \\
Br$\gamma$$^a$ & 5.4'' & 1.67 $\pm$ 0.41 &    $\sim$ 6.2 & Fischer et al.  1987 \\ \hline
\[[Fe II]$^b$  &       &  4.91 $\pm$ 0.32 &  $\sim$   10   & Forbes \& Ward 1993 \\
\hline \hline
Line & Aperture & EW$_{3D}$$^c$ & EW$_{lit}$ &Reference \\ 
 &  & [\AA] & [\AA] & \\ \hline
H$_2$ 1-0 S(1) & 3.4'' $\times$ 12.0'' & 5.82 $\pm$ 0.33 & 6.0 $\pm$ 0.5 & Vanzi et al. 1998\\
H$_2$ 1-0 S(0) & 3.4'' $\times$ 12.0'' & 1.06 $\pm$ 0.22 & 3.0 $\pm$ 1.0 & Vanzi et al. 1998\\
Br$\gamma$     & 3.4'' $\times$ 12.0'' & 3.30 $\pm$ 0.81 & 2.5 $\pm$ 0.5 & Vanzi et al. 1998\\
He I           & 3.4'' $\times$ 12.0'' & $\leq$ 1.0  &     1.1 $\pm$ 0.5 & Vanzi et al. 1998\\
\[[Fe II]        & 3.4'' $\times$ 12.0'' & 5.78 $\pm$ 0.38  & 3.5 $\pm$ 1.0  & Vanzi et al. 1998\\
\hline \hline
\end{tabular}
\end{center}
$^a$ Derived from the Pa$\beta$ flux of Ward et al. (1987) and assuming
Pa$\beta$/Br$\gamma$ $\approx$ 6.
\\
$^b$ Assuming [Fe II] 1.64 $\mu$m = 0.71 [Fe II] 1.24 $\mu$m using the
value of Ward et al. 1987.
\\
$^c$ Measured in a 4.6'' aperture.
\end{table}

\begin{table}[htb]
\caption{
\label{xxt07}
Fluxes and equivalent widths of the BLR, CLR and NLR emission lines in
NGC~3227}
\begin{center}
\begin{tabular}{lrrr}\hline \hline
Line& $\lambda_o$ & Flux& EW \\
       & [$\mu$m]  & [10$^{-17}$ W m$^{-2}$] & [\AA] \\ \hline 
\[[FeII]   & 1.53 & 1.53 $\pm$ 0.34 & 2.02 $\pm$  0.45   \\
\[[FeII]   & 1.64 & 4.80 $\pm$ 0.15 & 6.33 $\pm$  0.21   \\
\[[FeII]   & 1.74 & 2.03 $\pm$ 0.24 & 2.92 $\pm$  0.35   \\
\[[SiVI]$^a$   & 1.96 & 2.15 $\pm$ 0.17 & 2.46 $\pm$  0.20   \\
\[[AlIX]   & 2.04 & 0.75 $\pm$ 0.14 & 1.62 $\pm$  0.30   \\
Br$\gamma$ & 2.17 & 1.66 $\pm$ 0.26 & 3.66 $\pm$  0.57   \\   
\hline \hline
\end{tabular}
\end{center}
$^a$ possible contribution of the \htwo 1-0 S(3) are not taken into
account
\\
Fluxes and equivalent widths (EW) of the emission lines of the BLR,
CLR and NLR measured in a 3.6'' center on the nucleus of NGC~3227.
The quoted uncertainties for the line fluxes include the uncertainties
from the choise of the baseline for the continuum. The total absolute
uncertainty is larger, since the calibration uncertainties will be
added. The uncertainties for the equivalent widths were obtained from
the uncertainties of the line fluxes and the corresponding neighboring
continuum. The spatial resolution is 1.6'' ($H$ band) and 1.3'' ($K$
band). (10$^{-14}$ erg s$^{-1}$ cm$^{-2}$ = 10$^{-17}$ W m$^{-2}$)
\end{table}

\begin{table}[htb]
\caption{
\label{xxt08}
Fluxes and equivalent widths of the stellar absorption lines in
NGC~3227}
\begin{center}
\begin{tabular}{lrrrr}\hline \hline
Linie & $\lambda_o$ & Flux& EW \\
  & [$\mu$m] & [10$^{-17}$ W m$^{-2}$] & [\AA] \\ \hline
SiI   & 1.59 & -0.91 $\pm$ 0.12 & 1.23 $\pm$ 0.16 & \\ 
CO 6-3  & 1.62 & -1.42 $\pm$ 0.10 & 1.88 $\pm$ 0.13 & \\
CO 7-4  & 1.64 & -0.86 $\pm$ 0.09 & 1.13 $\pm$ 0.12 & \\
CO 8-5  & 1.66 & -0.90 $\pm$ 0.07 & 1.20 $\pm$ 0.09 & \\
CO 9-6  & 1.68 & -0.76 $\pm$ 0.14 & 1.02 $\pm$ 0.19 & \\
CO 10-7 & 1.71 & -0.94 $\pm$ 0.14 & 1.28 $\pm$ 0.19 & \\   
Mg/Al & 2.11 & -0.53 $\pm$ 0.19 & 1.15 $\pm$ 0.41 & \\
NaI   & 2.21 & -1.47 $\pm$ 0.19 & 3.27 $\pm$ 0.42 & \\
CaI   & 2.26 & -0.58 $\pm$ 0.17 & 1.35 $\pm$ 0.40 & \\
$^{12}$CO (2-0)  & 2.29 & -3.41 $\pm$  0.37 &  7.91 $\pm$  0.86 & \\
$^{12}$CO (3-1)  & 2.32 & -2.56 $\pm$  0.21 &  5.93 $\pm$  0.49 & \\
$^{12}$CO (4-2)  & 2.35 & -2.80 $\pm$  0.26 &  6.48 $\pm$  0.60 & \\
$^{12}$CO (5-3)  & 2.38 & -3.78 $\pm$  0.40 &  8.75 $\pm$  1.25 & \\
$\sum$ $^{12}$CO &      & -12.32 $\pm$ 1.19 & 28.53 $\pm$ 12.13 & \\
CO               &      & -22.24 $\pm$ 3.19 & 51.53 $\pm$ 58.71 & \\   \hline
$\frac{CO 6-3}{SiI}$         & & & & 1.53 $\pm$  0.23   \\
$\frac{CO 6-3}{CO2-0}$       & & & & 0.48 $\pm$  0.06   \\
log($\frac{CO 6-3}{SiI}$)    & & & & 0.18 $\pm$  0.07   \\
log($\frac{CO 6-3}{CO 2-0}$) & & & & -0.32 $\pm$ 0.06   \\   
\hline \hline
\end{tabular}
\end{center}
Fluxes and equivalent widths (EW) of the stellar absorption lines
measured in a 3.6'' center on the nucleus of NGC~3227.
The quoted uncertainties for the line fluxes include the uncertainties
from the choise of the baseline for the continuum. The total absolute
uncertainty is larger, since the calibration uncertainties will be
added. The uncertainties for the equivalent widths were obtained from
the uncertainties of the line fluxes and the corresponding neighboring
continuum. The spatial resolution is 1.6'' ($H$ band) and 1.3'' ($K$
band). (10$^{-14}$ erg s$^{-1}$ cm$^{-2}$ = 10$^{-17}$ W m$^{-2}$)
\end{table}

\begin{table}[htb]
\caption{
\label{xxt02}
NIR flux densities and colors of NGC~3227 at a resolution of 0.55''}
\begin{center}
\begin{tabular}{rrrrrrrrrr}\hline \hline
Aper. & ~~~J & & ~~~H & & ~~~K & & J - H & H - K & J - K \\
~[''] & [mJy] & [mag] & [mJy] & [mag] & [mJy] & [mag] & [mag] & [mag] & [mag] \\ \hline
 0.6 &    6.57& 13.58&  9.84& 12.59& 13.06& 11.71 & 0.98 & 0.88 & 1.87 \\
 1.0 &   14.63& 12.71& 21.19& 11.76& 26.61& 10.94 & 0.95 & 0.82 & 1.77 \\
 1.4 &   22.43& 12.24& 31.07& 11.34& 37.18& 10.57 & 0.90 & 0.77 & 1.67 \\
 1.8 &   29.45& 11.95& 39.31& 11.09& 45.39& 10.36 & 0.86 & 0.73 & 1.59 \\
 2.2 &   35.68& 11.74& 46.40& 10.91& 52.23& 10.21 & 0.83 & 0.70 & 1.53 \\
 2.6 &   41.06& 11.59& 52.49& 10.77& 58.02& 10.09 & 0.81 & 0.68 & 1.50 \\
 3.0 &   45.74& 11.47& 57.84& 10.67& 63.09& 10.00 & 0.80 & 0.67 & 1.47 \\
 3.6 &   51.91& 11.33& 64.84& 10.55& 69.65&  9.89 & 0.79 & 0.65 & 1.44 \\
 4.6 &   60.53& 11.17& 74.58& 10.39& 78.67&  9.76 & 0.77 & 0.63 & 1.40 \\
 6.0 &   70.51& 11.00& 85.71& 10.24& 89.07&  9.63 & 0.76 & 0.62 & 1.37 \\
 9.0 &   87.79& 10.76&105.05& 10.02&107.94&  9.42 & 0.74 & 0.60 & 1.34 \\
 9.1 &   88.27& 10.76&105.62& 10.02&108.51&  9.41 & 0.74 & 0.60 & 1.34 \\
\hline \hline
\end{tabular}
\end{center}
NIR flux densities and colors measured in different apertures center
on the nucleus of NGC~3227. The calibration uncertainties in each band for
the individual nights are 7.8 \% ($J$ band), 4.8 \% ($H$ band) and 5.7
\% ($K$ band). Additional uncertainties from the calibration of the
standard star 35 Leo as well as systematic uncertainties amount in
a total uncertainty for each band of about 10\%.
\end{table}

\begin{table}[htb]
\caption{
\label{xxt05}
Flux values of the northern knot}
\begin{center}
\begin{tabular}{crr}\hline \hline
Band & [mJy] & [mag]  \\ \hline
V & 0.07 & 19.37 \\
J & 0.33 & 16.82 \\
H & 0.42 & 16.02 \\
K & 0.39 & 15.52 \\
\hline \hline
\end{tabular}
\end{center}
Values are measured in a 1'' aperture centered in the middle of the
extended structure. The uncertainties correspond to the ones mentioned
in Table \ref{xxt02}. The uncertainties of the HST $V$ band are higher,
since only a differential calibration has been possible (see text).
\end{table}

\begin{table}[htb]
\caption{\label{xxt10}Star Formation Scenarios in the Nucleus}
\begin{center}
\begin{tabular}{lrrrrrr}\hline \hline
 & Obs. & RSG$^a$ & RSG$^b$ & RSG$^c$ & AGB$^d$ & RG$^e$  \\ \hline
\% M stars (H)                   & 30  & 38  & 33  & 40  & 27  & 27  \\
\% M stars (K)                   & 25  & 24  & 22  & 11  & 12  & 12  \\
\% stellar (H)                   & 65  & 65  & 65  & 65  & 65  & 65  \\
\% stellar (K)                   & 40  & 40  & 40  & 40  & 40  & 40  \\
L$_K$ [10$^8$ L$_{\odot}$]       & 1.4 & 1.4 & 1.4 & 1.4 & 1.4 & 1.4 \\
L$_{LyC}$ [10$^8$ L$_{\odot}$]   & 3.9$^1$ & 2.3 & 0.3 & 0.0 & 0.0 & 3.3 \\
L$_{bol}$ [10$^9$ L$_{\odot}$]   & 9.3 & 10.0& 12.5& 9.0 & 3.6 & 6.2 \\
$\nu_{SN}$ [10$^{-2}$ yr$^{-1}$] & 2.4 & 2.3 & 5.0 & 0.2 & 0.0 & 0.4 \\
m$_{st}$ [10$^8$ M$_{\odot}$]    & 11$^2$  & 0.3 & 0.9 & 1.5 & 3.7 & 5.6 \\
m$_{gas}$ [10$^8$ M$_{\odot}$]   & 0.9 & 0.4 & 1.2 & 2.4 & 8.2 & 26  \\
SFR [10$^{-2}$ yr$^{-1}$]        &     & 9.1 & 0.9 & 0.0 & 0.0 & 26.8\\
\hline \hline
\end{tabular}
\end{center}
$^1$ total observed Br$\gamma$ flux used
\\
$^2$ dynamical mass in the inner 3.6'' derived from the rotation curve
(Schinnerer, Eckart \& Tacconi 2000) 
\\
The observed values are derived assuming A$_V$=4$^{mag}$  
and using the following measured values in a
3.6'' aperture: S$_K$=27.9mJy, S$_{5GHz}$=24mJy,
F$_{Br\gamma}$=1.66$\times$10$^{-14}$ ergs cm$^{-2}$.
The equations can be found in section \ref{xxop}.
m$_{st}$ and m$_{gas}$ denote the present-day stellar mass and
the consumed gas mass to date, respectively.
\\
All starburst models are calculated using a Salpeter IMF with mass
cut-offs of 0.8 \solm ~to 120 \solm. 
\\
RSG$^a$: age of 15 Myr and decay time of 3 Myr 
\\
RSG$^b$: age of 25 Myr and decay time of 3 Myr 
\\
RSG$^c$: age of 50 Myr and decay time of 3 Myr 
\\
AGB$^d$: age of 0.5 Gyr and decay time of 3 Myr 
\\
RG$^e$: age of 10 Gyr and continuous star formation.
\end{table}

\begin{table}[htb]
\caption{
\label{xxt09}
Variation of the model parameters}
\begin{center}
\begin{tabular}{llrlrrr}\hline \hline
 & Variied &m$_{up}$ & & t$_{burst}$ & A$_V$ & dust in \% of \\
Fig. & parameter &[M$_{\odot}$] & kind&[10$^7$ yr] & [mag] & $K$ band flux \\ \hline
\ref{xxa12} (a), (e) & age & 120 & decaying & 0.7 & 0.0 & 0.0 \\
		     & age & 120 & decaying & 1.56 & 0.0 & 0.0 \\
		     & age & 120 & decaying & 7.0 & 0.0 & 0.0 \\
		     & age & 120 & constant & 700.0 & 0.0 & 0.0 \\ \hline
\ref{xxa12} (b), (f) & m$_{up}$ & 120 & decaying & 0.7 & 0.0 & 0.0 \\
		     & m$_{up}$ &  30 & decaying & 0.7 & 0.0 & 0.0 \\
		     & m$_{up}$ &  10 & decaying & 0.7 & 0.0 & 0.0 \\ \hline
\ref{xxa12} (c), (g) & A$_V$ & 120 & decaying & 1.56 & 0.0 & 0.0 \\
		     & A$_V$ & 120 & decaying & 1.45 & 5.0 & 0.0 \\
		     & A$_V$ & 120 & decaying & 1.56 &10.0 & 0.0 \\ \hline
\ref{xxa12} (d), (h) & dust & 120 & decaying & 1.56 & 0.0 & 0.0 \\
		     & dust & 120 & decaying & 1.56 & 0.0 & 10.0 \\
		     & dust & 120 & decaying & 1.56 & 0.0 & 25.0 \\
\hline \hline
\end{tabular}
\end{center}
Variied parameters in Figure \ref{xxa12} for $H$ and $K$ band spectra.
Age (t$_{burst}$), upper mass cut-off (m$_{up}$), kind of burst as
well as extinction (A$_V$) and contribution of hot dust emission (with
a temperature of 900~K) to the $K$ band flux were changed.
\end{table}

\clearpage

\end{document}